\documentclass[fleqn,10pt]{wlscirep}
\usepackage[utf8]{inputenc}
\usepackage[T1]{fontenc}

\usepackage{url}

\usepackage{amsmath, amssymb, graphicx}
\usepackage{varwidth}

\usepackage{subcaption}

\usepackage{comment}

\usepackage{csquotes} 

\usepackage{multirow}
\usepackage{ulem}
\usepackage{orcidlink}
\usepackage{afterpage}

\usepackage{hyperref}

\usepackage{caption}
\title{Comparing Probabilistic Influence-Spreading Centralities to Commonly-Used Centrality Measures in Directed and Weighted Networks}

\author[1,*]{Juuso Luhtala}
\author[1,*]{Vesa Kuikka \orcidlink{0000-0002-3677-816X}}
\author[1,*]{Kimmo K. Kaski \orcidlink{0000-0002-3805-9687}}

\affil[1]{Aalto University, Department of Computer Science, School of Science, P.O. Box 11000 00076 Aalto Finland}

\affil[*]{juuso.luhtala@aalto.fi, vesa.kuikka@aalto.fi (corresponding author), kimmo.kaski@aalto.fi}


\begin{abstract}
The Influence-Spreading Model (ISM) introduces three probabilistic centrality measures: out-centrality, in-centrality, and ISM betweenness centrality. Out-centrality measures the average probability that a node influences others, while in-centrality measures the average probability that others influence a node. ISM betweenness centrality measures the change in total probabilistic influence when a node is removed. These measures depend on edge transmission probabilities and allow walks up to a specified maximum length. We compare the ISM centrality measures to commonly used weighted variants of out-degree, in-degree, closeness, shortest-path betweenness, and Katz centrality in directed, weighted networks using four real-world online social networks and nine synthetic networks generated by Erdős-Rényi, navigable small-world, and directed scale-free models. For the synthetic networks, the edge probabilities are drawn from three beta distributions. We evaluate the similarity in centrality values and their ranking using Pearson correlation and Spearman's rank correlation coefficients. Results show strong correlations between the ISM out-centrality and weighted out-degree and outward Katz centrality, particularly for low edge probabilities. Conversely, relationships between the ISM in-centrality and other measures vary with network topology, sometimes yielding negative correlations. Correlations between the ISM betweenness and the shortest-path betweenness are also topology-dependent and weaken as alternative influence paths become more relevant. Overall, standard centrality measures can approximate the influence of broadcasting influence but often miss the nuances of receiving influence and probabilistic intermediary roles.
\end{abstract}

\begin{document}

\flushbottom
\maketitle

\thispagestyle{empty}

\section{Introduction} \label{Introduction}

In network science, various network centrality measures have been proposed over the years to quantify how important or influential nodes are in a network. 
These centrality measures are designed to measure specific types of information, material transmission, or flow characteristics in a network. Every centrality measure proposes a unique and different way of calculating the centrality of a node. However, despite these conceptual and methodological differences between centrality measures, multiple studies have found varying levels of positive correlation between them. One cause of positive correlations between centrality 
measures could be some similarities between their definitions, i.e. they measure the same thing at least partially. A more nuanced explanation for the positive correlations between centrality measures is that the underlying network structure causes higher-order correlations. Of course, there can be truth in both explanations at the same time.

To address the complexities of the importance of a node in networks, the Influence-Spreading Model (ISM) was introduced in \cite{kuikkaSciRep, kuikka2022efficiency}. This model offers a dynamic flow-based perspective to characterize network properties, treating edge weights as transmission probabilities that govern the propagation of physical, informational, or social entities in networks. By considering all possible walks up to a defined maximum length, the ISM allows us to analytically calculate the cumulative pairwise influence transmission probabilities between all network nodes. Building upon this foundation, the ISM framework intrinsically defines three distinct centrality measures: ISM out-centrality, quantifying a node's capacity to act as an influence source; ISM in-centrality, measuring a node's susceptibility to being influenced; and ISM betweenness centrality, evaluating a node's role as an intermediary in propagation. These metrics are naturally aligned with the underlying principles of the model. The influence-spreading model and its associated metrics are suitable for analyzing directed, weighted, and even disconnected networks, thus offering a highly adaptable alternative to classical centrality paradigms.

In this study, we compare the centrality measures defined by the influence-spreading model with other centrality measures in the context of directed and weighted networks. Degree, closeness, betweenness, and eigenvector centrality are widely recognized as foundational measures in the study of social networks \cite{borgatti2006, wasserman1994, brandes2005}, alongside extended measures such as Katz and PageRank centralities \cite{newman2010}. Here, our focus is on analyzing networks that are directed and weighted and which are not necessarily strongly or even weakly connected. The directed and weighted variants of degree, betweenness, and Katz centralities naturally function in networks that are not strongly or weakly connected, whereas for closeness centrality, we consider a generalized variant capable of handling such disconnected structures. Consequently, our analysis considers degree, closeness, betweenness, and Katz centrality as standard centrality measures, focusing on their directed and weighted variants. We exclude eigenvector centrality because it does not necessarily perform well on non-strongly connected networks, as explained in more detail in Section~\ref{katz_centrality}. In addition, we only consider centrality measures designed for general-purpose network analysis. This focus naturally rules out centralities like PageRank, since it was specifically tailored to rank web pages.

This paper continues our previous study \cite{luhtala2025} by extending the analysis to directed and weighted networks, in which edge probabilities are not restricted to a single constant value. The main goal is to clarify when commonly used directed and weighted centrality measures can be used as proxies for probabilistic influence and when explicitly probabilistic ISM centrality measures provide additional information. More specifically, we ask how the relationship between the ISM centrality measures and other network centrality measures depends on its structure, edge probability distributions, and direction of propagation of influence. We also expand the comparison to Katz centrality, since it accounts for all walks in a graph and serves as a natural comparison for ISM out-centrality and ISM in-centrality. The theoretical similarities and differences between Katz centrality and ISM out- and in-centrality are explored in more detail in Section \ref{out-in-comparison-katz}. An integral part of this study is the inclusion of real-world network datasets with empirically estimated edge probabilities, providing a realistic setting for comparing ISM centrality measures with other commonly used centrality measures. In addition, we create synthetic datasets with simulated edge probabilities to supplement the analysis.

This paper is organized as follows. In Section \ref{Related_work}, we review a wide variety of work related to our study, including comparisons of centrality measures and theoretical work on similarities between some other centrality measures. In Section \ref{standard_measures}, we introduce the weighted and directed variants of standard centrality measures, which we compare against the centrality measures defined by the influence-spreading model. In Section \ref{ISM_new_metrics}, the influence-spreading model is introduced, and the centrality measures it defines are explained in detail. In Section \ref{Data_methodology}, we introduce real-world networks with associated empirically estimated edge probabilities, explain the logic and reasoning behind synthetic network creation and edge probability simulation, and present the comparison methodology via the use of Pearson correlation and Spearman's rank correlation coefficients. In Section \ref{Results}, we review the results of centrality measure correlation comparisons and relate them to previously existing theory and empirical results. In Section \ref{discussion}, we discuss our findings and in Section \ref{conclusion} draw conclusions of our study.

\section{Related Work} \label{Related_work}

The relationship between the Influence-Spreading Model (ISM) and classical diffusion models, such as the Independent Cascade (IC) and Linear Threshold (LT) models \cite{chen2014}, is rooted in how influence propagates through a network. Traditional diffusion models often rely on Monte Carlo simulations to estimate the expected number of influenced nodes, which is a computationally intensive process that can be challenging for complex network structures. Although analytical solutions for dynamical phenomena in networks can be difficult to derive \cite{random_walk}, various approaches exist, including Markovian methods, matrix algebraic models, and mean-field approximations \cite{kuikkaSciRep}. These methods offer different ways to model network dynamics, often by averaging interactions or focusing on simplified state dependencies.

To overcome the computational and analytical limitations of traditional approaches, the ISM introduces a novel analytical framework to calculate the probabilities of directed influence between all pairs of nodes. It achieves this by considering all walks of finite length that are shorter than or equal to a predefined parameter value, $L_{max}$ \cite{kuikkaSciRep}. This parameter is not a fundamental limitation of the model, as the influence spreading process typically converges quickly. As a result, the contributions of very long walks to the total probability become negligible. This framework is related to the Influence Maximization (IM) problem, which aims to identify a seed set of nodes that maximizes the expected spread of influence. The ISM provides a computationally efficient method for quantifying the probabilistic reach of nodes in directed and weighted networks, without requiring exhaustive stochastic simulations.

Multiple studies have compared various centrality measures that are considered standard. In \cite{valente2008}, unweighted versions of degree, closeness, betweenness and eigenvector centrality were compared using relatively small networks collected from multiple sociometric studies. These measures were chosen for comparison because of their popularity among network analysts. The most relevant correlation results with respect to our study were that out-degree and outward closeness centrality had $0.81$ Pearson correlation, while the in-degree and inward closeness had $0.55$ Pearson correlation. In another larger study, seventeen different centrality measures, among them degree, closeness, shortest paths betweenness and Katz centrality, were compared using Spearman's rank correlation in a study involving 212 network datasets \cite{oldham2019}. The networks used in the study were not directed and either weighted or unweighted. Some networks used in the study were converted from a directed network to an undirected one by adding bi-directional edges. Additionally, only the largest connected component was analyzed if there were multiple connected components. The results of this study indicated that the rankings of centrality measures are generally positively correlated, but the strength of Spearman's rank correlation varies across networks. Additionally, centrality measure correlations in weighted networks were only marginally weaker than in their unweighted counterparts. In \cite{mohamadichamgavi2024}, both Pearson and Spearman's rank correlation coefficients were calculated between degree, closeness, shortest paths betweenness and eigenvector centrality in undirected and unweighted synthetic networks created by the Erd\H{o}s-R\'enyi, Barab\'asi-Albert and the Watts-Strogatz network models. The researchers found that correlations are positive, but especially in case of Watts-Strogatz and Barab\'asi-Albert networks, the strength of correlation depends heavily on model parameters.

Correlations between centrality measures can be caused by the underlying network structure. In \cite{schoch2017}, degree, closeness, shortest paths betweenness, and eigenvector centrality were compared using undirected networks with isolated nodes removed from them. The researchers found that if the structure of a network is close to a threshold graph, then the correlations between centrality measures are explained more by the network structure than by any similarities in the centrality measures. In a different study \cite{ricardo2015}, Pearson correlation coefficients were calculated  between the logarithmic values of multiple centrality measures in both real and synthetic undirected and unweighted networks. The compared centrality measures included degree, closeness, shortest paths betweenness, and eigenvector centrality. It was found that the correlations were generally higher on synthetic networks than in real networks. Synthetic networks were produced using either the Barab\'asi-Albert model or the Watts-Strogatz model. The results prompted the authors to propose that networks or network models might have a specific \emph{correlation profile}, i.e., that there is important information in the specific wiring pattern of each network. The correlation profile can cause higher average correlations between many centrality measures in a specific network compared to other networks.

The centrality measures based on walk counting use the notion that the entries $[A^k]_{ij}$ of an adjacency matrix $A$ record the number of walks of length $k$ from node $i$ to node $j$. In \cite{benzi2013}, various centrality measures based on walk counting were theoretically and empirically compared using undirected (and unweighted) synthetic networks created from the Barab\'asi-Albert and Watts-Strogatz models. Empirical comparisons used the Pearson correlation coefficient and the intersection distance to quantify the differences between the centrality measures. Among the comparisons, Katz centrality was compared to resolvent-based subgraph centrality, and similar comparisons were made between the row sums of the matrix exponential and exponential-based subgraph centrality. In \cite{benzi2015}, the relationships between multiple walk counting-based centrality measures, degree centrality, and eigenvector centrality were analyzed mathematically. The reason for selecting these centrality measures for comparison was that they can all be interpreted as functions of the adjacency matrix. The study considered all variants of networks, from undirected and unweighted to directed and weighted networks. For directed and weighted networks, they proved that if the network is strongly connected, then when the Katz centrality parameter $\alpha$ approaches $0$ from above, the rankings produced by outward and inward Katz centrality coincide with the rankings produced by weighted out-degree and in-degree, respectively. Similarly, when $\alpha$ approaches $1/\rho(A)$ from below, where $\rho(A)$ is the largest absolute eigenvalue of $A$, then the rankings produced by outward and inward Katz centrality coincide with right and left eigenvector centrality, respectively. It is important to note that this result is true only when the network is strongly connected. The authors of this study also performed computational comparisons between these centrality measures, which provided empirical support for the mathematical proofs.

In weighted centrality measure comparisons, edge weights can have a significant effect compared to unweighted centrality measures. For example, if a particular centrality measure is very sensitive to changes in edge weights, i.e. if a small change in edge weights causes significant ranking changes, then it has major implications for centrality measure comparisons on weighted networks. In \cite{segarra2015}, the stability of various weighted centrality measures is analyzed, and the results are supported by empirical experiments. According to the authors' definition of stability for weighted centrality measures, weighted degree, out-degree, in-degree, closeness and eigenvector centrality are found to be stable, but weighted shortest paths betweenness centrality is not found to be stable. A related question is how dissimilar networks become, when the same network structure is given two different sets of edge weights? In \cite{jiang2021}, the question of quantifying the dissimilarity of weighted networks was addressed. The authors developed a dissimilarity metric for weighted networks. Among their experiments, they compared the values of their proposed dissimilarity metric by simulating edge weights from different probability distributions.

Later we analyze centrality measure correlations on multiple real-world online social networks, where edge probabilities, i.e. influence probabilities between users, have been empirically estimated based on user activity and interactions between the users. This empirical estimation of influence probabilities can be complicated by many users of online social networks being mainly passive consumers of content with only infrequent active participation. The majority of users in online social networks have been theorized and empirically verified to be so-called \emph{lurkers}, i.e. users with very low-level, infrequent posting activity who mainly passively consume content on the platform \cite{gong2015, antelmi2019}. In \cite{gong2015}, a brief literature review reports that the percentage of lurkers in various online social networks ranges from 46 \% to 92 \%. The "90-9-1 rule" by Jakob Nielsen proposes that 90 \% of users are lurkers, 9 \% are moderately active users, and only 1 \% are active users \cite{nielsen2006}. In \cite{antelmi2019}, the "90-9-1 rule" was tested empirically against data collected from Twitter from August 2017 to December 2017. The researchers found that the empirical data supported a rule of 75-20-5, i.e that approximately 75 \% of users are lurkers. Although the estimates for the percentage of lurkers vary, we can conclude that the majority of users in online social networks tend to be lurkers. The lurker phenomenon complicates empirical edge probability estimation for online social networks since, for many users, there is very little data to base the edge probability estimates on. The very low activity of lurkers results in user action logs that contain only a few observations, and interaction history between users might be lacking altogether. Missing interaction history between users that are otherwise in a followee-follower relationship requires the assignment of baseline influence probabilities.

\section{Network Centrality Measures} \label{standard_measures}

In this section, we introduce the directed and weighted variants of the centrality measures we consider "standard". First, we introduce some important terminology that is frequently used in this paper. We do not allow selfloops in networks, i.e., edges from a node to itself are not allowed. We also do not allow multiedges, i.e., multiple edges in the same direction. We frequently use the terms \emph{walk} and \emph{path}. A walk in a network is an alternating sequence of nodes and edges, where both nodes and edges can be repeated without restriction, as long as the walk respects the direction of the edges. A path is a walk in which nodes and edges cannot be repeated. A network is strongly connected if there exists a directed path from any node to all other nodes. A network is weakly connected if there exists an undirected path from any node to all other nodes when directed edges are replaced with undirected ones. In this paper, we define a directed network $G = (V, E)$ such that a directed edge from node $u \in V$ to node $v \in V$ is denoted by $e = (u, v) \in E$. All the edge weights we consider in this work are edge probabilities, with one exception being shortest paths calculations, when we transform the edge probabilities into distances using a logarithm-based transformation, which is explained in Section \ref{local_shortest_measures} in more detail. We denote edge probabilities with $p(e) \in (0, 1]$ for all $e \in E$. A missing edge is considered synonymous with an edge probability of zero.

We introduce the standard centrality measures in the order in which the centrality measures make more and more use of the global characteristics of the network they measure. The directed variants of degree centrality, in-degree and out-degree, are only based on local information of the network. Inward and outward closeness centrality are based on calculating shortest paths, where the distance and number of shortest paths can differ based on direction. As the name implies, shortest paths betweenness centrality uses information about the shortest paths in a network. We refer to shortest paths betweenness centrality, also as standard betweenness centrality. Katz centrality is based on accounting for all possible walks in a network, with longer walks being penalized with a special decay/damping parameter. All standard centrality measure values are calculated using the library \texttt{NetworkX} \cite{hagberg2008}.

In directed networks, a distinction between \emph{broadcasting} and \emph{receiving} centrality measures can be made \cite{benzi2015}. Out-degree and outward Katz centrality are considered broadcasting centrality measures, and in-degree and inward Katz centrality are considered receiving centrality measures. A similar distinction can be made with outward and inward closeness centrality, respectively. Furthermore, ISM out-centrality and ISM in-centrality fit naturally in this framework as broadcasting and receiving centrality measures, respectively. Shortest paths betweenness centrality and ISM betweenness centrality are outside of this type of classification, since the shortest path betweenness considers both directions, and ISM betweenness measures how much the sum of total influence probabilities changes when a node is removed from a network.

\subsection{Centrality Measures Based on Local Information and Shortest Paths} \label{local_shortest_measures}

Let us consider a directed and weighted network $G = (V, E)$. For a node $v \in V$, out-degree measures the number of outgoing edges that are incident to $v$ and in-degree measures the number of incoming edges that are incident to $v$ \cite{brandes2005}. Weighted out-degree measures the sum of the weights of outgoing edges that are incident to node $v$, and weighted in-degree measures the sum of the weights of incoming edges that are incident to node $v$ \cite{brandes2005}. If edge weights are probabilities that quantify the strength of the influence relationship between nodes, then the weighted out-degree represents how much a node influences its neighbors in aggregate, and the weighted in-degree represents how much the neighbors of a node influence it in aggregate. We note that it is possible that a node can have a large value of out-degree or in-degree, but the weighted versions can have comparatively much smaller values. This would represent a situation where a node has many (outgoing or incoming) connections, but the strength of those connections is weak. Both unweighted and weighted versions of out-degree and in-degree are centrality measures that are based only on local information of the network in the neighborhood of node $v$.

As explained before, edge probabilities are defined as a function $p: E \rightarrow (0, 1]$ for all $e \in E$. When edge weights are probabilities, and we wish to use weighted centrality measures that are based on shortest paths, then in the absence of additional distance information, there needs to be a systematic way of transforming probabilities to distances. One such way of transforming probabilities to distances was presented in the Appendix of \cite{luhtala2025}. The interpretation presented in the article is that finding the most probable path is transformed into a problem of finding the shortest path in a network. Mathematically, if $e_1, \ldots, e_m \in E$, with $m = 1, \ldots, |E|$, are edges in a path, maximizing the product $\prod_{k=1}^m p(e_k)$ is equivalent with maximizing the sum $\sum_{k=1}^m \log p(e_k)$ which is equivalent with minimizing $-\sum_{k=1}^m \log p(e_k)$ which is equivalent with minimizing $\sum_{k=1}^m \log (1 / p(e_k))$. With the transformation $\log (1 / p(e))$ for each edge $e \in E$, we can interpret $\log (1 / p(e))$ as a non-negative distance value. This distance grows when edge probability becomes smaller, and the distance shrinks when edge probability becomes greater. We exclude the possibility of $p(e) = 0$ for obvious reasons. Although we allow $p(e) = 1$ in our edge probability definition, it produces a distance value of $\log (1 / 1) = 0$, which can cause problems in shortest path calculations. Therefore, in practice an edge probability of exactly or nearly one could be replaced with a slightly smaller probability value that does not cause numerical problems. With the logarithmic transformation, it becomes possible to use algorithms, such as Dijkstra's algorithm, to find shortest paths in a network. In the absence of other distance information, this transformation gives a natural interpretation for the shortest path as the most probable path. Therefore, the information contained in edge probabilities can be used in a natural and systematic way. In \cite{halasz2026}, edge probabilities were transformed into distances in a similar way.

For undirected networks, closeness centrality measures the inverse of the average distance between a node and all other nodes \cite{wasserman1994}, When the network is directed, a distinction needs to be made between inward distance, which measures the distance to a node based on incoming paths, and outward distance, which measures the distance to a node based on outgoing paths \cite{hagberg2008}. When a network has multiple connected components (undirected case) or strongly connected components (directed case), the question becomes how to define closeness centrality in this case? In this case, Wasserman and Faust define closeness centrality by considering the influence range of a node and by making closeness centrality proportional to how many nodes are reachable compared to the size of the whole network \cite{wasserman1994}. Based on these notions, we define outward closeness centrality $c_{outC}(v)$ and inward closeness centrality $c_{inC}(v)$ of a node $v$ as
\begin{equation}
c_{outC}(v) = \frac{J_v^{out}}{N - 1} \frac{J_v^{out}} {\sum_{u \in J_v^{out}} d(v, u)} \quad \text{and} \quad c_{inC}(v) = \frac{J_v^{in}}{N - 1} \frac{J_v^{in}} {\sum_{u \in J_v^{in}} d(u, v)},
\end{equation} where $J_v^{out}$ is the number of reachable nodes for node $v$ using outward paths, $J_v^{in}$ is the number of nodes that can reach node $v$ using inward paths, $N$ is the total number of nodes in the network, $d(v, u)$ is the distance measured from node $v$ to node $u$ and $d(u, v)$ is the distance measured from node $u$ to node $v$. In both $J_v^{out}$ and $J_v^{in}$ node $v$ is excluded from the calculation. For directed networks, we note that in general $d(v, u) \neq d(u, v)$. Distance can be any non-negative real number when considering edge weights, or it can be interpreted as one when not considering edge weights. The factors $J_v^{out} / (N - 1)$ and $J_v^{in} / (N - 1)$ ensure that the closeness centrality value calculated from a limited set of reachable nodes is proportional to the total size of the network. If a node is totally isolated, then it is given a closeness centrality value of zero. In \cite{brandes2005}, there is a discussion on the drawbacks of the definition of closeness centrality that is proportional to how many nodes are reachable compared to the size of the whole network.

Standard betweenness centrality is based on how often a node lies on the shortest paths between two other nodes. For this reason, it is also called shortest-path betweenness centrality. Standard betweenness centrality $c_B(v)$ for node $v$ is
\begin{equation}
c_B(v) = \sum_{\substack{s \neq v \neq t \\s \neq t}}\frac{\sigma_{st}(v)}{\sigma_{st}},
\end{equation} where $\sigma_{st}$ denotes the number of shortest paths between nodes $s$ and $t$ and $\sigma_{st}(v)$ denotes the number of shortest paths between $s$ and $t$ that contain the node $v$ \cite{brandes2005}. Notably, for directed networks, the counts of shortest paths can differ by direction, but this definition takes direction into account \cite{hagberg2008}. Shortest-path calculations can be based either on non-negative distances between nodes (the weighted case) or on all distances being one between nodes (the unweighted case).

\subsection{Katz Centrality: A Centrality Measure that Accounts For All Possible Walks in a Network} \label{katz_centrality}

For a directed graph $G = (V, E)$ with associated weighted or unweighted adjacency matrix $A$, the outward Katz centrality $c_{outK}(i)$ and the inward Katz centrality $c_{inK}(i)$ for a node $i \in V$ are defined as
\begin{equation}
c_{outK}(i) = \sum_{j=1}^n \sum_{k=0}^{\infty} \alpha^k [A^k]_{ij} = \sum_{j=1}^n [(I - \alpha A)^{-1}]_{ij} \quad \text{and} \quad
c_{inK}(i) = \sum_{j=1}^n \sum_{k=0}^{\infty} \alpha^k [A^k]_{ji} = \sum_{j=1}^n [(I - \alpha A^{T})^{-1}]_{ij},
\end{equation} where $n$ denotes the number of nodes, $I$ is the identity matrix, $\alpha$ is called the decay/damping parameter and the $\alpha$ parameter needs to satisfy the condition $0 < \alpha < 1 / \rho(A)$, where $\rho(A)$ is the largest absolute (modulus) eigenvalue of $A$ \cite{brandes2005, noferini2024}. The upper limit $1 / \rho(A)$ for the damping factor $\alpha$ ensures that the infinite sum converges to a finite value and that the matrices are invertible. When $0 < \alpha < 1 / \rho(A)$, the infinite sum converges to the closed form solution $\sum_{k=0}^{\infty} \alpha^k (A^k) = (I - \alpha A)^{-1}$. We also note that $[A^k]_{ji} = [(A^{k})^{T}]_{ij} = [(A^{T})^k]_{ij}$ for all $i = 1, \dots, n$, $j = 1, \dots, n$ and $k \in \mathbb{N}$. Therefore, the outward and inward Katz centrality can also be expressed with the closed form solution involving the inverse matrices.

Outward and inward Katz centrality correspond to the row sum and column sum of the inverse matrix $(I - \alpha A)^{-1}$, respectively. Outward and inward Katz centrality describe the hub and authority score of a node, respectively \cite{benzi2015}. It is notable that for undirected graphs, there is no distinction between inward and outward Katz centrality, since the adjacency matrix is always symmetric. For an unweighted adjacency matrix $A$, $(A^k)_{ji}$ tells the number of walks of length $k$ between nodes $j$ and $i$. When the adjacency matrix $A$ consists of positive real-valued weights, $(A^k)_{ji}$ tells the weighted sum of walks of length $k$ from node $j$ to node $i$ \cite{arrigo2024}. It is also worth noting that we start counting walk lengths from zero, which results in the term $A^0 = I$, whereas some other definitions of Katz centrality start counting walk lengths from one. These two methods produce no difference in the centrality rankings, since the same constant value is being added to every centrality score from the $A^0 = I$ term.

In \cite{benzi2015}, the authors report that Katz centrality is most informative when the $\alpha$ parameter is of the form $\tau / \rho(A)$, where $\tau \in [0.5, 0.9]$. Based on numerical experiments, they report that for $\alpha < 0.5 / \rho(A)$, the rankings of outward Katz centrality are very close to the out-degree, and for $\alpha > 0.9 / \rho(A)$, the rankings of outward Katz centrality are very close to the right eigenvector centrality. In \cite{aprahamian2016}, the authors researched a way of selecting the $\alpha$ parameter such that it would closely match the rankings produced when using the matrix exponential $e^A$ instead of the matrix resolvent $(I - \alpha A)^{-1}$. In order to avoid further complexity, we select the $\alpha$ parameter simply with the choices $\tau \in \{ 0.1, 0.5, 0.9 \}$. This selection gives enough variety for the $\alpha$ parameter while keeping experimental complexity under control.

It is useful to emphasize the information that is contained in the powers of the adjacency matrix $A^k$. If the adjacency matrix $A$ is unweighted, then elements of $A^k$ tell how many walks of length $k$ are between the nodes. For example, if there are five walks of length three from node $i$ to node $j$, then $(A^3)_{ij} = 5$. If the edge weights are probabilities, then the probability of each individual walk is added up, and the entry $(A^3)_{ij}$ tells the sum of the individual walk probabilities. A singular walk probability is the product of the edge probabilities that exist on that particular walk, where the interpretation is that each edge probability represents an independent event and the product of these edge probabilities represents the joint probability of these independent events \cite{arrigo2024}.

Katz centrality is closely related to eigenvector centrality, as both of them are based on the principle that a node is considered important if it is connected to other important nodes \cite{newman2010}. This foundational idea represents their primary similarity. In strongly connected networks, Katz centrality captures structural characteristics that are highly similar to those of eigenvector centrality \cite{bonacich2001}. In fact, as the damping parameter $\alpha$ approaches its upper limit of $1 / \rho(A)$, the rankings produced by outward Katz centrality converge with those of right eigenvector centrality \cite{benzi2015}. However, the key difference between the two measures lies in their mathematical formulation and how they address networks that are not strongly connected. In networks with disconnected components, eigenvector centrality can propagate zero centrality values to other nodes, even when it is undesirable\cite{newman2010, bonacich2001}. Katz centrality mitigates this limitation by assigning a constant baseline centrality to every node (originating from the $A^0 = I$ term), which prevents the centrality scores of weakly connected nodes from collapsing to zero \cite{newman2010}.

Given that our analysis includes both real-world and synthetic networks that are not strongly connected, we exclude eigenvector centrality from our comparisons in favor of the more structurally robust Katz centrality. Finally, while Newman's interpretation of Katz centrality is recursive and mirrors the conceptual foundation of eigenvector centrality \cite{newman2010}, this study focuses on its combinatorial interpretation based on walk counting. Despite these different perspectives, both interpretations describe the same underlying centrality measure \cite{borgatti2006b}.

\section{The Centrality Measures Defined by the Influence-Spreading Model} \label{ISM_new_metrics}

The influence-spreading model \cite{kuikkaSciRep, kuikka2022efficiency} is designed to describe probabilistic influence spreading between nodes in a network. The influence-spreading model can describe two types of spreading: Either complex contagion or simple contagion. In complex contagion walks, from a source node to a target node, are almost unrestricted, with the only restrictions being that the influence probability calculation ends when the target node is reached and that the walk length is limited by the maximum allowed walk length. Outside of these two conditions, cycles and self-intersecting walks are allowed with no restrictions placed on them. In contrast, simple contagion focuses on self-avoiding walks. In this paper, concentrate exclusively only on the complex contagion version of the influence-spreading model.

The influence-spreading model and its calculation of influence probabilities have been previously described in detail in \cite{kuikkaSciRep, kuikka2022efficiency, luhtala2025}. Therefore, in this paper, we describe the model only briefly while focusing on the key ideas and definitions. The key contribution of the influence-spreading model is to quantify a node's influence on other nodes as a probability. This is achieved by considering edge weights as edge probabilities and using this information in conjunction with the rules of probability. At the heart of this process is how the probabilities of two walks are combined.

The probability of a walk is the product of the edge probabilities that exist on that particular walk. In other words, each transmission of influence between nodes is treated as an independent event, and the joint probability of these transmission events is the product of the edge probabilities. In the influence-spreading model, walks are denoted by sequences of nodes and nodes are numbered starting from one. The probabilities of two walks $\mathcal{L}_1$ and $\mathcal{L}_2$ are combined by using their longest common prefix (LCP) walk $\mathcal{L}_3$ and the rules of probability. The influence-spreading model assumes that two walks $\mathcal{L}_1$ and $\mathcal{L}_2$ are conditionally independent given a common prefix walk $\mathcal{L}_3$ i.e.\ $\mathbb{P}(\mathcal{L}_1 \cap \mathcal{L}_2 \mid \mathcal{L}_3) = \mathbb{P}(\mathcal{L}_1 \mid \mathcal{L}_3) \mathbb{P}(\mathcal{L}_2 \mid \mathcal{L}_3)$, where $\mathbb{P}$ denotes probability. The complete logic of the combination process of walk probabilities is described in detail in \cite{luhtala2025}. The formula for combining the probabilities of the walks $\mathcal{L}_1$ and $\mathcal{L}_2$ given a common prefix walk $\mathcal{L}_3$ is as follows
\begin{align}
\mathbb{P}(\mathcal{L}_1 \cup \mathcal{L}_2) = \mathbb{P}(\mathcal{L}_1) + \mathbb{P}(\mathcal{L}_2) - \frac{\mathbb{P}(\mathcal{L}_1) \mathbb{P}(\mathcal{L}_2)}{\mathbb{P}(\mathcal{L}_3)}. \label{path_combination}
\end{align} The derivation of this formula is explained in more detail in \cite{luhtala2025}.

We next describe the logic of the walk combination process. All walks of length less than or equal to $L_{max}$ between a source node and a target node are explored in the influence-spreading model. The walks are then sorted in lexicographic order, where shorter walks come first and walks with the same length are sorted according to where they first differ. Once all walks are sorted in lexicographic order, the walks are combined starting from the longest common prefix walks and using formula (\ref{path_combination}). Progressively shorter LCP walks are considered until all walk probabilities have been combined into one probability. Walks that have the same LCP length can be combined in any order. Finally, we note that in practice, a more effective algorithm is used, but both versions produce the same exact output. This more effective algorithm does not require enumerating all possible walks, and it was described in detail in \cite{kuikka2022efficiency}.

Once the influence probability is calculated between all node pairs, the influence-spreading model has finished its computation. The output of the influence-spreading model \cite{kuikkaSciRep, kuikka2022efficiency} is called the influence spreading matrix. The elements of this matrix are the probabilities of influence spreading from a source node $i$ to a target node $j$ for all pairs of $N \times N$ nodes within a network consisting of $N$ nodes. Given an influence spreading matrix $M$, out-centrality $C_{out}$ and in-centrality $C_{in}$ are defined as follows
\begin{equation}
    C_{out}(i)=\frac{1}{N-1}\sum_{j \in V \setminus \{ i \}} M(i,j) \quad \text{and} \quad C_{in}(j)=\frac{1}{N-1}\sum_{i \in V \setminus \{ j \}} M(i,j), \quad i=1,...N.
\end{equation}
We see that out-centrality values correspond closely with the row sums of the influence spreading matrix, while in-centrality values correspond closely with the column sums of the influence spreading matrix. We assume that the influence spreading process begins with probability one from the initial node. Since a node's influence on itself is defined as always being one, we ignore the diagonal in the definitions of the out-centrality and the in-centrality. The values of in-centrality and out-centrality are always between zero and one, since the row and column sums are normalized by the factor $1/(N-1)$. Out-centrality measures how much a node influences other nodes on average, and in-centrality measures how much a node is influenced by other nodes on average. Out-centrality can be considered to be a broadcasting centrality measure, whereas in-centrality can be considered to be a receiving centrality measure.

The influence-spreading model also defines its own betweenness centrality measure, called ISM betweenness centrality or ISM betweenness. ISM betweenness centrality is based on how much the sum of all influence probabilities being transmitted in a network changes when a node and its incident edges are removed from the network. ISM betweenness centrality $b_{ISM}(v)$ of a node $v$ is defined as
\begin{equation} \label{eq:ISM_betweenness}
b_{ISM}(v) = \frac{\mathcal{C} - \mathcal{C}_v}{\mathcal{C}}, \quad \text{where} \quad \mathcal{C} = \sum_{\substack{s, e \in V \\ s \neq e}} M(s, e), \quad \mathcal{C}_v = \sum_{\substack{s, e \in V \setminus \{ v \} \\ s \neq e}} M^{*}(s, e),
\end{equation} and $M^{*}$ is the influence spreading matrix calculated from the network without node $v$ and its incident edges.

In this paper, we use the value $L_{max} = 20$, since this same parameter value was used in the previous study comparing ISM centrality measures to standard centrality measures in undirected networks \cite{luhtala2025}. The effect of the $L_{max}$ parameter is that larger values allow longer walks and thus more probabilistic influence can go through a network. However, at some point, influence probability values start to converge since longer walks carry less weight due to probabilities being multiplied with each other. The full specification of the influence-spreading model \cite{kuikkaSciRep} also describes a time-dependent probability component that approaches one when time grows unbounded. As in the previous study \cite{luhtala2025}, we assume in this paper that the time-dependent probability component is always one.

\subsection{Out- and In-Centrality in Comparison to Weighted Katz Centrality: Similarities and Differences} \label{out-in-comparison-katz}

Katz centrality considers all walks of infinite length, with the alpha parameter acting as a damping factor which ensures convergence to a finite value. In ISM out- and in-centrality, all walks up to a maximum allowed walk length are considered. In the influence-spreading model, the calculation between a source and a target node stops when the target node is reached. This nuance slightly restricts possible walks, while Katz centrality truly considers every single possible walk with absolutely no restrictions.

For Katz centrality in weighted networks, where edge weights are probabilities, the weight of an individual walk is the product of the edge probabilities, i.e. the weight of a walk is the joint probability of the independent edge probabilities that exist on that walk. Walk probabilities are simply added together, and the alpha parameter ensures convergence. Therefore, Katz centrality does not have an in-built probability interpretation due to probabilities being simply added together with no consideration for joint probabilities of multiple walks.

In the context of the influence-spreading model, walk probabilities are joined by considering the probability of their common walk via the addition rule for probabilities. Centrality measures have an in-built probability interpretation. If the joining of walk probabilities is removed from the influence-spreading model, then the values produced for finite sums are almost the same as Katz centrality if the damping parameter alpha is removed and only walks with a maximum allowed walk length are considered.

To further illustrate the similarities and differences between Katz centrality and the calculation of the influence spreading matrix, let us consider the following example. Let $A$ be the unweighted adjacency matrix of network $G = (V, E)$ and let $W$ be the weighted adjacency matrix of $G$, where the weights are probabilities. If $(A^k)_{ji} = 5$, then there are five walks of length $k$ from node $j$ to node $i$. Let us denote these five walks with $\mathcal{L}_1, \dots, \mathcal{L}_5$. The corresponding entry of the weighted adjacency matrix $W$, will be $(W^k)_{ji} = \mathbb{P}(\mathcal{L}_1) + \dots + \mathbb{P}(\mathcal{L}_5)$ i.e. the entry $(W^k)_{ji}$ is the sum of the individual walk probabilities. An individual walk probability $\mathbb{P}(\mathcal{L}_h)$, for $h = 1, \dots, 5$, is the product of the edge probabilities that exist on that particular walk.

The difference in the calculation of the influence spreading matrix is that the influence-spreading model considers the common walk via the use of the addition rule for probabilities before the walk probabilities are joined in a union. If the influence-spreading model ignored the common walk between two walks altogether and simply added the probabilities of the walks, then for finite maximum walk lengths the influence-spreading model would produce almost the same exact values than Katz centrality, if Katz centrality was only calculated for finite maximum walk lengths and the damping factor alpha was set to one (since the damping factor alpha is not needed for finite sums). In our centrality measure comparisons, we normalize the values of Katz centrality to facilitate a more natural comparison between ISM centrality measures defined on the $[0, 1]$ interval.

\section{Data and Methodology} \label{Data_methodology}

We define a directed influence network $G = (V, E)$ to be such that an influence relationship from node $u \in V$ to node $v \in V$ is denoted by a directed edge $e = (u, v) \in E$. The intensity of this relationship is quantified with the influence probability $p(e) \in (0, 1]$ for all $e \in E$. We note that when $u$ influences $v$, it implies that $(u, v) \in E$ and that there is a follower relationship, where $v$ follows $u$. We note that in a strictly following network, the same information could be coded by reversing the direction of the edges in the influence network. However, our interest is in influence probability and how much a node influences other nodes and therefore, the direction $u \rightarrow v$ accurately describes a situation when $u$ influences $v$.

We will use both real and synthetic influence network data to perform our comparisons. Real influence network data with empirically estimated edge influence probabilities consists of four different publicly available online social networks, where an action log of the users and the interaction history between them were used to estimate directed influence probabilities between them. We have performed an extensive search of publicly available network data \emph{with associated empirically estimated edge probabilities}, but we have found this type of dataset to be rare. In \cite{chen2014}, it is commented that finding publicly available data with both a social network and action log of users is relatively rare. Since suitable real-world data is hard to find, we supplement our comparison by creating directed synthetic networks with simulated edge probabilities. Additionally, simulated edge probabilities give the possibility of studying centrality measure correlations under the effect of different edge probability distributions on the same network structures. The synthetic networks are created from three different network models, which allow for directed networks. The edge probabilities are randomly generated from different probability distributions. 

\subsection{Online Social Network Datasets with Empirically Estimated Edge Probabilities} \label{real_networks_empirical_edge_probabilities}

The four different publicly available online social network datasets that have associated empirical edge probability estimates consist of three datasets collected from Twitter and one collected from the Russian social networking site VK (formerly known as VKontakte). The three Twitter network datasets have been collected when the social network and microblogging platform now known as X was still known as Twitter. In order to accurately reflect when these datasets were collected, we always refer to them as Twitter datasets even though the social network is now named differently.

The first and second Twitter datasets were collected by the same research group in Australia \cite{zhang2019}. The first Twitter dataset was location-based in Darwin, Australia, and it was collected from November 2017 to December 2017. The second Twitter dataset was event-based during the thoroughbred horse race competition Melbourne Cup 2017. The third Twitter dataset consists of the Twitter network formed by members of the 117th US Congress and the tweets made by the members from February 2022 to June 2022 \cite{fink2023}. The fourth dataset has been collected from the Russian social network VK \cite{logins2019}. The authors of this article do not specify the exact collection time, but we estimate that the dataset was collected sometime before or around the year 2019, based on when the original article and its associated code and data repository were published. All four datasets are available in code and data repositories linked in the original articles where the datasets were originally used or published. The dataset published by the Australian research group \cite{zhang2019} required additional data analysis and programming to get the empirical edge probabilities, but thankfully, the researchers had published helpful code to make this possible. The two other datasets were provided already in a usable format.

The authors of the US Congress Twitter dataset say that, to their knowledge, the dataset they have published is the only publicly available Twitter dataset that empirically estimates pairwise influence probabilities between nodes \cite{fink2023}. However, we have managed to find two additional publicly available Twitter datasets that perform similar empirical edge probability estimation. Additionally, in the VK social network data, we found a fourth publicly available dataset that also has associated estimated edge probability data. To our knowledge, these four network datasets with realistic and empirically estimated edge probabilities have never been used in the same study to compare weighted centrality measures and the rankings they produce. Since for all four directed network datasets the edge probabilities have been empirically estimated from data, we have the most realistic situation to perform centrality measure comparisons.

Network statistics for the four publicly available datasets are collected in Table \ref{table:network_statistics}. We note that these statistics do not use edge weight information and, therefore, average shortest path lengths for the largest strongly connected component are calculated with all edge distances being assumed to have an equal value. For all networks except the Darwin Twitter network, the size of the largest strongly connected component is close to the size of the whole network. Only the US Congress Twitter network is weakly connected. All networks display quite high reciprocity i.e. the number of edges pointing to both directions compared to the total number of edges. This statistic is consistent with earlier findings in \cite{mislove2007}, where it was found that online social networks are characterized by high levels of reciprocity.

When an action log of users or interaction history between users is available, it is possible to use this data to estimate empirical edge probabilities. If edge probabilities are assumed to be static, then when a user $u$ tries to influence user $v$ this event can be viewed as a Bernoulli trial and the maximum likelihood estimate for the influence probability $p(u, v)$ is then
\begin{equation} \label{edge_prob_estimate}
p(u, v) = \frac{A_{u2v}}{A_u},
\end{equation} where $A_{u2v}$ denotes the number of $u$'s actions that have influenced $v$ and $A_u$ denotes the total number of actions performed by $u$ \cite{goyal2010}. Additionally, there are other approaches for estimating the static edge probability \cite{chen2014}. One of them is the Jaccard index, where $A_{u|v}$ is used instead of $A_u$ in the denominator and where $A_{u|v}$ denotes the number of $u$'s or $v$'s combined actions. An even more refined approach is the so-called credit model, where each neighbor of $v$ gets only partial credit for influencing $v$. In all four real-world network datasets, the empirical estimation of edge probabilities was based on the maximum likelihood estimate in formula (\ref{edge_prob_estimate}). However, there were also some small additional refinements between the approaches.

The Darwin, Australia-based Twitter dataset and the Twitter dataset related to Melbourne Cup 2017 were both collected by the same research group, and the same methodology was used to build the directed influence network and to estimate the empirical influence probabilities between users in the influence network \cite{zhang2019}. We argue that the estimation done by the Australian research group was the most sophisticated. In their methodology, when a user (node) $v$ reacts to user $u$'s message, this was recorded as user $u$ having an influence on user $v$. The possible  Twitter reactions were divided into five categories: 1) $v$ retweets $u$'s message, 2) $v$ likes $u$'s message, 3) $v$ replies to $u$'s message, 4) $v$ quotes $u$'s message and 5) $v$ who is a follower of $u$, posts similar content to $u$ within a certified time limit. Furthermore, influence was further divided into explicit and implicit influence, and their probabilities were estimated separately. Tweeting, quoting and replying were defined to be explicit influence, while retweeting and liking were considered to be implicit influence. The explicit influence probability $p^e(u, v)$ and implicit influence probability $p^i(u, v)$ were estimated with a similar formula to (\ref{edge_prob_estimate}), which emphasized the total number of explicit or implicit actions and how many of these had an influence on $v$. Implicit influence was given only half the weight of explicit influence. Additionally, to scale down the probability estimate when there are too few observed actions $A_u$ (either explicit or implicit), they multiplied the raw estimate by the function $f(A_u) = 1 / (1 + \exp (1 - A_u))$. The final influence probability estimate was $p(u, v) = 1 - (1 - p_1)(1 - p^e(u, v))(1 - p^i(u, v))$, where $p_1 = 0.001$ is assumed to be a baseline influence probability from the follower $u$ towards the follower $v$. When an interaction history between users $u$ and $v$ was missing, but there existed a follower relationship between them, the edge probability was assigned the baseline value $p_1 = 0.001$. We note that the vast majority of the edge probabilities in the Darwin and Melbourne Cup 17 Twitter datasets consist of baseline influence probabilities, since for most edges there was no action log or interaction data available in order to estimate influence probabilities.

For the Twitter dataset from the US Congress, user $v$'s actions of retweeting, quoting, replying, or mentioning $u$'s post were denoted as influence of user $u$ towards user $v$ \cite{fink2023}. Then the influence probability was straightforwardly estimated with equation (\ref{edge_prob_estimate}). The influence probabilities in the VK social network dataset were calculated based on similarities between the wall post, where it is assumed that user $u$ causes user $v$ to post similar content on their wall, thus influencing $v$. The estimation of influence probability seemed to follow the formula (\ref{edge_prob_estimate}), but this is not crystal clear, as there are some notational inconsistencies in the original article. However, the distribution of edge probabilities is, broadly speaking, similar to the Australian Twitter datasets. Therefore, we assume that these notational inconsistencies did not significantly impact the data analysis and probability estimation.

\begin{figure} [ht]
\includegraphics[width=\textwidth]{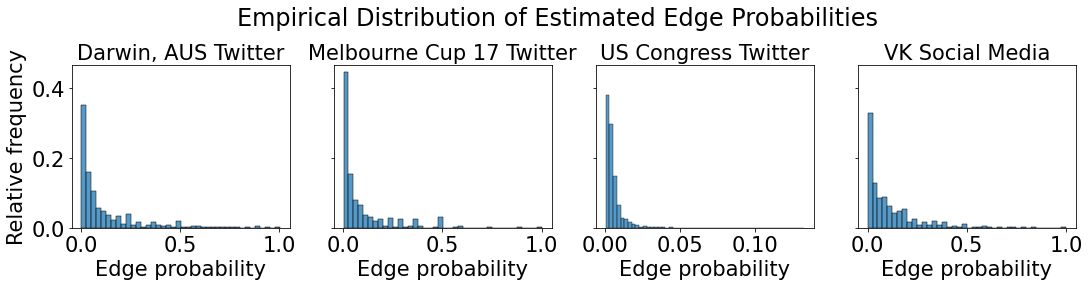} 
\caption{Empirical distribution of edge probabilities for the four network datasets. Edge probabilities were estimated from interaction history between users. For Darwin and Melbourne Cup 2017, Twitter influence networks only estimated probabilities are reported, i.e. baseline values of influence (probability = 0.001) are excluded.} 
\label{fig:distribution_edge_probabilities}
\end{figure}

Three of the four empirical edge probability distributions are similar to each other, with the only exception being the US Congress dataset, where the influence probabilities are much smaller than in the other datasets, and the maximum influence probability is only approximately $0.13$ whereas in other datasets the maximum is nearly or exactly $1.0$. The empirical distribution of edge probabilities for each of these networks is visualized in Figure \ref{fig:distribution_edge_probabilities} and the summary statistics of these distributions are available in the supplemental material. We note that the number of estimated edge probabilities is much smaller for the Darwin and Melbourne Cup 17 Twitter datasets than for the other two networks. The Darwin and Melbourne Cup 17 Twitter datasets have only 2080 and 7403 edges, respectively, with empirical edge probability estimates. The low amount of empirically estimated edge probabilities is consistent with the lurker phenomenon discussed earlier. To our knowledge, the US Congress Twitter and VK social network datasets did not use baseline probabilities, and all the edge probabilities found in the datasets are actual empirical estimates. The discrepancy in the number of estimated edge probabilities could be explained in multiple ways. Presumably, US Congress members have more motivation to actively participate and voice their opinions on Twitter than regular members of the public found in the other two Twitter datasets. In our understanding, the VK social network is similar to Facebook, which makes it more of a personal and friend-based social media with possibly more interactions between closer friends compared to more opinion-based social media like Twitter. Therefore, it would be understandable if users are more motivated to actively participate in a more friend-based social network compared to Twitter, where participation might expose one's opinions to more public criticism than in a more friend-based social media like VK.

\subsection{Synthetic Directed Networks with Simulated Edge Probabilities} \label{synthetic_networks_simulated_probabilities}

\subsubsection*{Synthetic Directed Networks} \label{synthetic_networks}

Since it is challenging to find real-world network data with associated empirically estimated edge probabilities, we supplement our comparisons with synthetic network data and with simulated edge probabilities. Before we assess the issue of how edge probabilities should be simulated, we discuss what type of synthetic networks we should generate. One central application area of the influence-spreading model is social networks, but the influence-spreading model can also be applied to other types of networks. Therefore, we do not need to place too many restrictions on how synthetic networks are generated, but we also should keep in mind social networks as a key application area of the influence-spreading model.

As social networks are a key application area, we should focus on the key characteristics. In \cite{broido2019}, many different types of network were studied to determine whether they can be described as scale-free networks. It was found that approximately 50\% of social networks showed no evidence of being scale-free. The remaining half of the social networks analyzed showed evidence of only weakly scale-free behavior. Other possible characteristics of real-world social networks are the small-world properties of short average path lengths and a high clustering coefficient, and there exists a long history of study of this phenomenon \cite{wasserman1994}. In \cite{mislove2007}, online social networks were analyzed and found evidence of scale-free and small-world properties.

The three network models which are used to produce directed synthetic networks are the directed Erd\H{o}s-R\'enyi model (DER) \cite{erdos1959, gilbert1959}, the navigable small-world model (NSW) \cite{kleinberg2000} and the directed scale-free network model (DSF) \cite{bollobas2003}. These three models were chosen since they capture a wide variety of possible network characteristics, and they are designed to generate directed networks. The basic formulations of the Watts-Strogatz small-world model and the Barab\'asi-Albert scale-free model describe undirected networks \cite{kleinberg2000, bollobas2003}. Therefore, we sought directed network model alternatives for creating small-world and scale-free networks. We note that probably none of these three network models fully capture the characteristics of online or real-world social networks. Therefore, the study of networks produced by these network models should be considered as a more general study of correlations between ISM centrality measures and standard centrality measures in networks having differing characteristics.

The directed Erd\H{o}s-R\'enyi model produces networks by creating a directed edge between two nodes with probability $P$. The random choice to connect two nodes is considered to be independent of other random choices. This model is very simple and does not necessarily correspond very well to real-world networks, but it allows one to easily manipulate the strong and weak connectivity of a network by varying the edge creation probability $P$.

The navigable small-world model was proposed not only to produce small-world networks but also to explain how the small-world phenomenon emerges in a network \cite{kleinberg2000}. The default navigable small-world model uses a two-dimensional $D \times D$ grid as the basis of the network, where $D \in \mathbb{N}_+$ and $\mathbb{N}_+$ denotes the positive integers. In the basic formulation of the navigable small-world model, there are $|V| = D^2$ nodes in the network $G = (V, E)$. Local directed contacts are formed between nodes based on the lattice distance $d_l((i, j), (k, l)) = |i - k| + |j - l|$ being less than or equal to $f \in \mathbb{N}_+$. Based on parameters $g \in \mathbb{N} = \mathbb{N}_+ \cup \{ 0 \}$ and $h \geq 0$, the model forms additional directed long-range contacts from node $u \in V$ to $g$ other nodes with probability proportional to $(d_l(u, v))^{-h}$, where $v \in V$. Since every node has a bi-directional relationship with at least its closest neighbors (when $f = 1$), the resulting network is always strongly connected. The navigable small-world network has a natural interpretation as an offline social network where neighbors always influence each other (they have a bi-directional relationship), and additionally, there are directed long-range influence relationships between non-neighbors that are not necessarily reciprocated.

 The directed scale-free model grows a network by always adding a single edge and possibly also adding a new node at each discrete time step \cite{bollobas2003}. Five parameters govern this process. The parameters $\omega > 0$, $\beta > 0$ and $\gamma > 0$ are interpreted as probabilities with the added constraint of $\omega + \beta + \gamma = 1$. The parameters $\delta_{\text{in}} \geq 0$ and $\delta_{\text{out}} \geq 0$ are constants which are added to the in-degree $d^-(v)$ and to the out-degree $d^+(v)$ of a node $v$, respectively, when choosing source and target nodes for new edges in the network. With probability $\omega$, a new node $v$ is connected to an existing node $u$, i.e. a directed edge $(v, u)$ is added to the network, where $u$ is chosen with a probability that is proportional to $d^-(u) + \delta_{\text{in}}$. With probability $\beta$ a new edge is added from node $u$ to node $v$, where existing nodes $u$ and $v$ are chosen independently with probability that is proportional to $d^+(u) + \delta_{\text{out}}$ and $d^-(v) + \delta_{\text{in}}$, respectively. With probability $\gamma$, an existing node $u$ is connected to a new node $v$, i.e. a directed edge $(u, v)$ is added to the network, where $u$ is chosen with a probability proportional to $d^+(u) + \delta_{\text{out}}$. The standard formulation of the directed scale-free model allows loops and multi-edges, but we remove all self-loops and multi-edges before analysis.
 
 The directed scale-free model was originally developed to accurately model web graphs, where the important nodes are those nodes that have many incoming edges, i.e. they are being linked to by many web pages \cite{bollobas2003}. However, in influence networks, the most influential nodes are those who have many outgoing edges. The directed scale-free model can also model an influence network, since, with careful parameter selection, we can emphasize outgoing edges. When modeling influence networks, we emphasize the value of $\gamma$ at the cost of diminishing  the value of $\omega$, since it is plausible to think that more often influence relationships tend to form from an existing node towards a new node in a network than vice versa. The directed scale-free model can possibly describe an influence network that operates on online social networks, where there are hub influencers with many outgoing edges (and thus many followers being influenced).

We generate three network datasets from each of the three stochastic network models $\text{DER}(P)$, $\text{NSW}(f, g, h)$ and $\text{DSF}(\omega, \beta, \gamma, \delta_{\text{in}}, \delta_{\text{out}})$ with suitably varied parameter values that emphasize the key characteristics of the networks the models are capable of producing. For directed Erd\H{o}s-R\'enyi networks, the key characteristic is to vary the connectivity of the network via parameter $P$. For navigable small-world networks, the key characteristic is to vary the overall reciprocity and the average shortest path length of the network by adding more long-range connections (parameter $g$) and by altering the probability of long-range connections being selected (parameter $h$). For directed scale-free networks, the key characteristic to vary is the size of the giant component of the network (mainly affected by parameter $\beta$) and reciprocity. In total, we created nine different synthetic networks with three networks produced from each network model. The parameters and the corresponding abbreviations for the nine synthetic networks are $P = 0.0055$ (DER1), $P = 0.015$ (DER2), $P = 0.05$ (DER3), $f, g, h = 1, 4, 2$ (NSW1), $f, g, h = 1, 20, 2$ (NSW2), $f, g, h = 1, 30, 0.5$ (NSW3), $\omega, \beta, \gamma, \delta_{\text{in}}, \delta_{\text{out}} = 0.01, 0.92, 0.07, 10, 10$ (DSF1), $\omega, \beta, \gamma, \delta_{\text{in}}, \delta_{\text{out}} = 0.001, 0.98, 0.019, 10, 10$ (DSF2) and $\omega, \beta, \gamma, \delta_{\text{in}}, \delta_{\text{out}} = 0.001, 0.99, 0.009, 25, 25$ (DSF3).  Overall, the parameter variations capture a wide range of varying connectivity through increasing number of edges and changing reciprocity values. Other parameter variations were also considered and they are available in the supplemental material. In Table \ref{table:network_statistics}, the synthetic networks are characterized by their network statistics.

\begin{table}
\vspace{-6pt}
\caption{Network statistics for real-world online social networks and synthetic networks. Edge weights are not considered when calculating basic network statistics. The average shortest path length is calculated for the largest strongly connected component (LSCC) in each network.}
\begin{tabular}{|p{2.4cm}|p{0.7cm}|p{0.8cm}|p{0.7cm}|p{0.7cm}|p{0.7cm}|p{0.7cm}|p{0.7cm}|p{0.7cm}|p{0.7cm}|p{0.7cm}|p{0.7cm}|p{0.7cm}|p{0.7cm}|}
 \hline
 \multicolumn{1}{|c|}{} & \multicolumn{4}{|c|}{Online Social Network} &
 \multicolumn{9}{|c|}{Synthetic Network} \\
 \hline
 Network statistic & Dar-win & Mel-Cup17 & Cong-ress & VK & DER-1 & DER-2 & DER-3 & NSW-1 & NSW-2 & NSW-3 & DSF-1 & DSF-2 & DSF-3 \\
 \hline
 Number of nodes & 3883 & 1437 & 475 & 2094 & 1000 & 1000 & 1000 & 1024 & 1024 & 1024 & 1000 & 1000 & 1000 \\
 Number of edges & 58799 & 72710 & 13289 & 24751 & 5476 & 15031 & 50451 & 7899 & 18260 & 33762 & 10666 & 19700 & 35800 \\
 LSCC \% nodes & 58.7 & 91.4 & 98.7 & 88.4 & 99.2 & 100.0 & 100.0 & 100.0 & 100.0 & 100.0 & 83.6 & 88.3 & 95.9 \\
 LSCC \% edges & 92.5 & 98.2 & 99.2 & 98.4 & 99.2 & 100.0 & 100.0 & 100.0 & 100.0 & 100.0 & 97.1 & 98.7 & 99.8 \\
 Weakly conn. & No & No & Yes & No & Yes & Yes & Yes & Yes & Yes & Yes & Yes & Yes & Yes \\
 Strongly conn. & No & No & No & No & No & Yes & Yes & Yes & Yes & Yes & No & No & No \\
 Reciprocity \% & 41.9 & 47.8 & 46.2 & 75.8 & 0.7 & 1.4 & 5.0 & 52.0 & 29.7 & 14.7 & 13.5 & 32.5 & 37.1 \\
 Average shortest path in LSCC & 3.2 & 2.7 & 2.4 & 3.3 & 4.3 & 2.8 & 2.0 & 4.2 & 2.9 & 2.3 & 2.9 & 2.3 & 2.1 \\
 \hline
\end{tabular}
\label{table:network_statistics}
\end{table}

\subsubsection*{Edge Probability Simulation} \label{edge_probability_simulation}

While there are plenty of publicly available directed network datasets, the same is not true for associated edge probability data. Therefore, in many applications where edge probabilities are needed, such as influence maximization, many simplifications have been used in the literature to obtain usable edge probabilities. Common approaches to produce edge probabilities are to set a constant value for all of them (uniform model), to use the so-called trivalency model, where a relatively low, medium or high probability value is drawn from a small set of values (usually three) with uniform random probability, or to make edge probabilities dependent on the in-degree of the target node (weighted cascade model) \cite{goyal2011, chen2014}. In \cite{halasz2026}, edge probabilities were alternatively generated from the uniform distribution on the interval $(0.05, 0.15)$, from the normal distribution with mean $0.1$ and standard deviation $0.05$ or from exponential distributions with different parameter values. To conform to the rules of probability, both the normal distribution and exponential distribution values were truncated on the interval $(0, 1)$.

In contrast to the aforementioned approaches, we generate edge probabilities from the beta distribution with suitably varied parameter values. The beta distribution is suitable for producing random probabilities, since, depending on its exact definition, it produces values either in the closed interval $[0, 1]$ or the open interval $(0, 1)$. The shape of the beta distribution can be changed depending on the parameters. This allows far greater flexibility compared to the exponential or normal distribution, since these distributions essentially maintain their shape even when their parameter values are changed. The definition using open interval $(0, 1)$ is suitable for producing edge probabilities, since we interpret an edge probability of zero as the edge being missing, and an edge probability of one can cause problems in shortest path calculations, since $\log 1/1 = 0$ produces a zero distance between nodes.

The beta distribution is characterized by two parameters $a > 0$ and $b > 0$ and denoted by $\text{Beta}(a, b)$. The probability density function $f(x | a, b)$ of the $\text{Beta}(a ,b)$ distribution is given by $f(x | a, b) = (1 / B(a, b)) x^{a - 1} (1 - x)^{b - 1}$, where $0 < x < 1$ and $B(a, b) = \Gamma(a) \Gamma(b) / \Gamma(a + b)$ is the Beta function that is defined with the help of the Gamma function $\Gamma(x) = \int_0^{\infty} e^{-t} t^{x - 1} dt$, for $x > 0$ \cite{krishnamoorthy2016}. The shape of the density function can be manipulated via its two parameters to be unimodal, continuously uniform with infinite number of modes or to be without a mode altogether. When the beta distribution has no mode, then its density has either a U-shape or a J-shape \cite{krishnamoorthy2016}. Similarly, the expected value, variance and other moments of the distribution depend on the parameters $a$ and $b$, thus allowing for great control and variability in the generation of random probabilities. We also note that the continuous uniform distribution on the interval $(0, 1)$ is a special case of the beta distribution, namely $\text{Beta}(1, 1)$. Therefore, the beta distribution is a flexible distribution that allows for many variations in the generation of random probability values.

We produce independent and identically distributed (iid) edge probability values from three different beta distributions: From $\text{Beta}(0.5, 4.5)$, $\text{Beta}(2, 8)$ and $\text{Beta}(1, 7/3)$. The expected values of the distributions are $0.1$, $0.2$ and $0.3$, respectively. The probability density and cumulative distribution functions of these beta distributions are illustrated in the supplemental material, and their theoretical statistics are also summarized there. The choice of the $\text{Beta}(0.5, 4.5)$ distribution parameters has been influenced by the empirical edge probability data found in the real-world online networks introduced in Section \ref{real_networks_empirical_edge_probabilities}. Maximum likelihood estimates for three of the four networks (Darwin Twitter, Melbourne Cup 17 Twitter and VK social network) are close to the chosen values of $a = 0.5$ and $b = 4.5$. The choice of the other two beta distributions is motivated by providing variety in the shape of the distribution ($\text{Beta}(2, 8)$) and by providing a much larger dispersion in edge probability values ($\text{Beta}(1, 7/3)$). In all selected beta distributions, there is an emphasis on lower influence probabilities on average. Lower influence probabilities, on average, can be thought of as being more realistic in influence networks between people.

\subsection{Comparison Methodology} \label{comparison_methodology}

We quantify our centrality measure comparisons by using Pearson correlation and Spearman's rank correlation coefficients \cite{kaptein2022}. The sample Pearson correlation coefficient is also called the product-moment correlation. Spearman's rank correlation coefficient is the product-moment correlation coefficient applied to the ranks of observations. When there are ties in the rankings, Spearman's rank correlation assigns the average rank of the ranking positions if no ties occurred. Puth et. al. \cite{puth2015} note that Spearman's rank correlation works well when there are ties in the data. Ties in rankings are a common situation when centrality measures are calculated. Additionally, we have also examined scatter plots of the centrality measure comparisons and selected the most illustrative examples to visualize in this paper.

Pearson correlation measures the linear association between two variables, whereas Spearman's rank correlation measures monotonic association between two variables \cite{puth2015}. Therefore, Spearman's rank correlation can capture a more complex relationship between two variables as long as there is a monotonic association between them. We use both correlation measures since the values of ISM centrality measures absolutely matter in applications, but we are also interested in possible rank correlations. Additionally, the previous study comparing ISM centrality measures to standard centrality measures \cite{luhtala2025} used both correlation coefficients. Puth et. al. \cite{puth2015} note that it is an appropriate application to use certain correlation coefficients if it gives an easier comparison to an earlier study.

\section{Results} \label{Results}

In this section, we present the correlation analysis between the centrality measures defined by the Influence-Spreading Model (ISM) and commonly used centrality measures. First, Section~\ref{subsec:baseline_findings} outlines the findings from our prior study on undirected, unweighted networks to establish a baseline. Section~\ref{results_real_networks} then evaluates the correlations on real-world online social networks using empirically derived edge probabilities, analyzing both the full networks and their largest strongly connected components (LSCC). We find that only centrality measure correlations involving the generalized version of closeness centrality change significantly when restricted to the LSCC. Finally, Section~\ref{results_synthetic_networks} analyzes results on synthetic directed networks across various simulated edge-weight distributions, noting that correlation results vary considerably more in synthetic networks than in empirical real-world networks.

\subsection{Baseline Findings from Undirected Networks} \label{subsec:baseline_findings}

To contextualize our new findings on directed and weighted networks, we briefly outline the baseline results obtained in our previous work~\cite{luhtala2025}. In that study, the centrality measures defined by the ISM (i.e. $C_{out}$, $C_{in}$, and $b_{ISM}$) were compared to unweighted degree, closeness, and shortest-paths betweenness centrality using undirected synthetic networks generated by Erd\H{o}s-R\'enyi, Watts-Strogatz, and Barab\'asi-Albert models. Edge probabilities in those experiments were held at a constant, uniform value across all edges, which was systematically varied to analyze its effect.

Several key properties of the ISM centralities emerged from this undirected baseline analysis. Both Pearson and Spearman's rank correlations were high between ISM betweenness ($b_{ISM}$) and shortest-path betweenness when edge probabilities were low. However, this correlation weakened significantly as edge probabilities increased, reflecting the entry of multi-walk alternatives. In the case of out-centrality monotonicity, the Spearman's rank correlation was exceptionally strong across all edge probabilities between ISM out-centrality ($C_{out}$) and degree centrality. In contrast, the initially high Pearson correlation weakened as edge probabilities rose, indicating a transition to a non-linear relationship. Similar behavior was observed when comparing $C_{out}$ with closeness centrality.

This earlier work highlighted that uniform edge probabilities and undirected structures produce predictable correlations. The current study expands on this by testing whether these baseline observations hold when we relax these assumptions and introduce directed edges, diverse topologies, and non-uniform, empirically or statistically derived edge weight distributions.

\subsection{Results for Online Social Networks with Empirically Estimated Edge Probabilities} \label{results_real_networks}

\begin{figure}
\includegraphics[width=\textwidth]{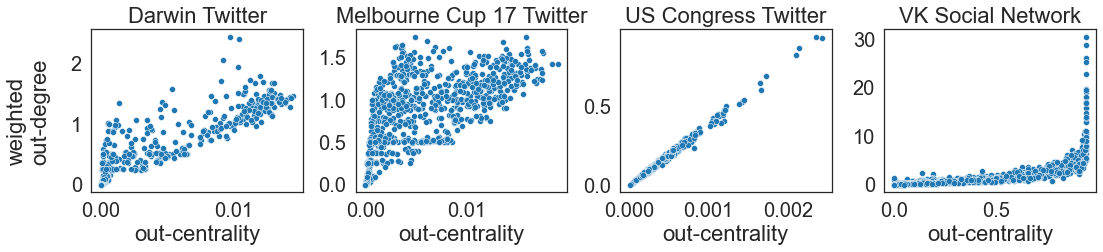} 
\caption{ISM out-centrality and weighted out-degree values for the four online social networks.}\label{fig:out-centrality_weighted_out-degree}
\end{figure}

For the online social networks with associated empirically estimated edge probabilities, we present the Pearson correlation and Spearman's rank correlation coefficients between ISM centrality measures and standard centrality measures in Table \ref{table:pearson_spearman_in_out_standard}. Across all studied networks, there is a moderate-to-strong positive Pearson correlation between out-centrality and weighted out-degree. Spearman's rank correlation between these measures is very high for all four networks. This relationship is visualized in Figure \ref{fig:out-centrality_weighted_out-degree}. For the US Congress Twitter network, the relationship between out-centrality and weighted out-degree is almost perfectly linear whereas the VK social network displays a non-linear yet monotonic-looking relationship. In contrast, there is more dispersion in the Darwin and Melbourne Cup 17 Twitter datasets. One way of explaining this is that, on average, influence probabilities in the US Congress datasets are very small, and thus very little influence can go through a network. Therefore, it is natural that out-centrality has nearly a linear relationship with weighted out-degree. It is also notable that the full range of out-centrality values is achieved in the VK social network dataset, whereas in the case of other three datasets, out-centrality values are quite low. In the VK social network, the largest values of weighted out-degree are also much higher than for the other three datasets.

\begin{figure}
\includegraphics[width=\textwidth]{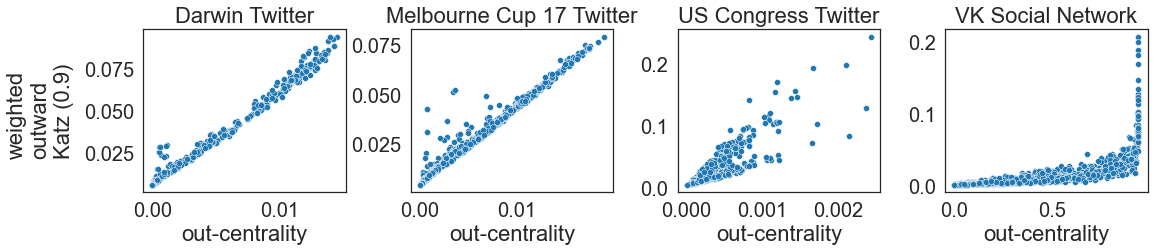} 
\caption{ISM out-centrality and (normalized) weighted outward Katz centrality (with $\alpha$ parameter equaling $0.9$ multiplied by the maximum allowed value) for the four online social networks.}\label{fig:out-centrality_weighted_outwards_Katz_09}
\end{figure}

There is a moderate-to-strong positive Pearson correlation and a very strong Spearman's rank correlation between out-centrality and weighted outward Katz centrality as anticipated based on theoretical similarities between these two centrality measures. In Figure \ref{fig:out-centrality_weighted_outwards_Katz_09}, the values of out-centrality and weighted outward Katz centrality for alpha multiplier $0.9$ are visualized. In the supplementary material, a scatter plot is shown between out-centrality and weighted outward Katz centrality with alpha multiplier $0.1$, and it looks almost identical to Figure \ref{fig:out-centrality_weighted_out-degree}, showing the relationship between the out-centrality and weighted out-degree. Supplemental material also shows that the Pearson correlation between weighted out-degree and weighted outward Katz centrality ($0.1$) is nearly perfectly positive, and in general, Pearson and Spearman's rank correlation between weighted out-degree and outward Katz centrality is quite strong for all tested values of the $\alpha$ parameter. This result is not unexpected based on the theoretical relationship between the weighted out-degree and the weighted outward Katz centrality \cite{benzi2015} as explained in Section \ref{Related_work}.

The relationship between the out-centrality and weighted outward closeness centrality displays a somewhat varying positive Pearson correlation. The relationship is visualized in the supplemental material. More detailed analysis on the strongly connected components reveals a much higher positive Pearson correlation for the Darwin and Melbourne Cup 17 Twitter datasets. The relatively small size of the largest strongly connected component of the Darwin Twitter dataset has a clear effect on the Pearson correlation, since analysis on the largest strongly connected component reveals a higher Pearson correlation ($0.651$). Additionally, if one singular outlier is removed from the Melbourne Cup 17 dataset, the Pearson correlation increases from $0.080$ to $0.651$. This outlier is caused by a single node having one outgoing edge $e$ with edge probability close to one. The resulting closeness centrality value calculated with the distance $\log (1 / p(e))$ results in a much larger closeness centrality value than exists for any other nodes in the network. This outlier has a great effect on the Pearson correlation, but it has only a negligible effect on Spearman's rank correlation. The largest strongly connected component of the Melbourne Cup 17 Twitter network also has a higher Pearson correlation of $0.789$. The Pearson correlations for the other two networks are much higher. For all four networks, Spearman's rank correlation between out-centrality and weighted outward closeness is quite strongly positive.

\begin{figure}
\includegraphics[width=\textwidth]{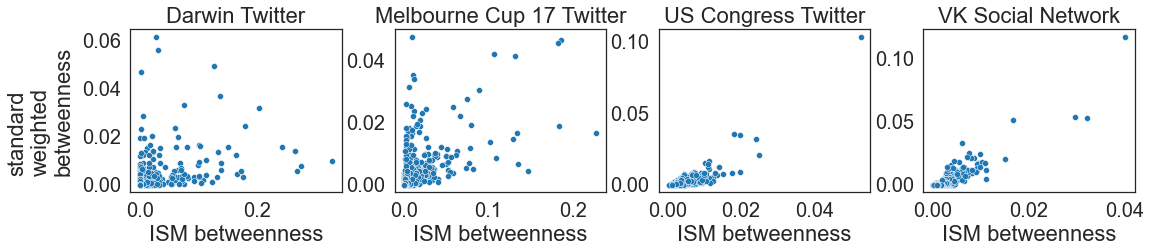} 
\caption{ISM betweenness centrality and weighted standard betweenness centrality for the four online social networks.}\label{fig:ISM_btw_standard_btw}
\end{figure}

In Figure \ref{fig:ISM_btw_standard_btw}, the relationship between the ISM betweenness centrality and the standard betweenness centrality is visualized. For the Twitter network of the US Congress and the VK social network, the association between these centrality measures seems more like a linear relationship. However, for the Darwin and Melbourne Cup 17 Twitter networks, there is no clear linear relationship. In Table \ref{table:pearson_spearman_in_out_standard}, we see that Pearson and Spearman's rank correlations have a range from moderately positive to strongly positive values. The definitions of these two betweenness centrality measures are quite different, and the correlation results support the notion that these centrality measures measure different things.

The relationship between in-centrality and inward standard centrality measures is more complicated than the relationship between out-centrality and most outward standard centrality measures. The relationship between out-centrality and weighted out-degree and outward Katz centrality displays a somewhat stable positive correlation. In Table \ref{table:pearson_spearman_in_out_standard}, we show that the in-centrality has varying levels of positive Pearson correlation with weighted in-degree and weighted inward Katz centrality. In \cite{luhtala2025}, it was recognized for undirected networks that when the edge probability increases for all edges, the in-centrality rankings tend to reverse. This means that when more influence is going through a network, the rankings produced by in-centrality do not remain stable. Instead, the least influenced nodes tend to become the most influenced nodes as more influence passes through a network. We hypothesize that this phenomenon is behind the centrality correlations between in-centrality and weighted in-degree and inward Katz centrality measures vary more strongly between different networks. The Pearson correlation between in-centrality and weighted inward closeness centrality varies quite a lot both for the full networks and also for the largest strongly connected component (see supplemental material). In Table \ref{table:pearson_spearman_in_out_standard}, Spearman's rank correlation values are higher than Pearson correlations, but a similar overall pattern between in-centrality and inward standard centrality measures can be seen. The relationship between in-centrality and out-centrality can display stronger positive Pearson correlation or it can be very weak. However, Spearman's rank correlations between in-centrality and out-centrality are never very strong for the four online social networks. All in-centrality and inward standard centrality measure values are visualized in the Supplemental material.

\begin{table}
\caption{Pearson correlation and Spearman's rank correlation coefficients between the centrality measures defined by the influence-spreading model and standard centrality measures. The alpha parameter of Katz centrality is a different multiple (in brackets) of the maximum allowed alpha parameter. *To avoid numerical problems in closeness centrality calculations, edge probabilities exceeding the value $0.9999$ have been assigned the value $0.9999$.}
\begin{tabular}{|p{6.1cm}|p{2.0cm}|p{1.4cm}|p{1.4cm}|p{1.4cm}|p{1.4cm}|}
 \hline
 \multicolumn{2}{|c|}{} &
 \multicolumn{4}{|c|}{Network Dataset} \\
 \hline
 Centrality comparison & Correlation coefficient & Twitter Darwin, Australia & Twitter MelCup 17 & Twitter US Congress & VK Social Network \\
 \hline
 \multirow{2}{\linewidth}{out-centrality VS weighted out-degree} & Pearson & 0.930 & 0.768 & 0.996 & 0.679  \\ 
 & Spearman & 0.959 & 0.921 & 0.996 & 0.956 \\
 \hline
 \multirow{2}{\linewidth}{out-centrality VS wtd. outward closeness} & Pearson & 0.392 & 0.080* & 0.808 & 0.723  \\
 & Spearman & 0.980 & 0.937 & 0.832 & 0.963 \\
 \hline
 \multirow{2}{\linewidth}{ISM betweenness VS wtd. std. betweenness} & Pearson & 0.470 & 0.590 & 0.831 & 0.893 \\
 & Spearman & 0.628 & 0.784 & 0.753 & 0.840 \\
 \hline
 \multirow{2}{\linewidth}{in-centrality VS weighted in-degree} & Pearson & 0.821 & 0.819 & 0.997 & 0.084 \\
 & Spearman & 0.919 & 0.794 & 0.997 & 0.545 \\
 \hline
 \multirow{2}{\linewidth}{in-centrality VS weighted inward closeness} & Pearson & 0.350 & 0.024* & 0.663 & 0.782 \\
 & Spearman & 0.981 & 0.868 & 0.925 & 0.601 \\
 \hline
 \multirow{2}{\linewidth}{in-centrality VS out-centrality} & Pearson & 0.804 & 0.537 & 0.159 & 0.246 \\
 & Spearman & 0.598 & 0.595 & 0.113 & 0.104 \\
 \hline
 \multirow{2}{\linewidth}{out-centrality VS outward Katz (0.9)} & Pearson & 0.996 & 0.990 & 0.838 & 0.709 \\
 & Spearman & 0.998 & 0.996 & 0.871 & 0.967 \\
 \hline
 \multirow{2}{\linewidth}{out-centrality VS outward Katz (0.5)} & Pearson & 0.970 & 0.892 & 0.987 & 0.721 \\
 & Spearman & 0.985 & 0.970 & 0.988 & 0.983 \\
 \hline
 \multirow{2}{\linewidth}{out-centrality VS outward Katz (0.1)} & Pearson & 0.939 & 0.791 & 0.999 & 0.692 \\
 & Spearman & 0.977 & 0.933 & 0.999 & 0.967 \\
 \hline
 \multirow{2}{\linewidth}{in-centrality VS inward Katz (0.9)} & Pearson & 0.829 & 0.910 & 0.884 & 0.092 \\
 & Spearman & 0.992 & 0.983 & 0.880 & 0.582 \\
 \hline
 \multirow{2}{\linewidth}{in-centrality VS inward Katz (0.5)} & Pearson & 0.846 & 0.902 & 0.991 & 0.092 \\
 & Spearman & 0.964 & 0.888 & 0.990 & 0.569\\
 \hline
 \multirow{2}{\linewidth}{in-centrality VS inward Katz (0.1)} & Pearson & 0.829 & 0.836 & 0.999 & 0.086 \\
 & Spearman & 0.947 & 0.813 & 0.999 & 0.552 \\
 \hline
\end{tabular}
\label{table:pearson_spearman_in_out_standard}
\end{table}

\subsection{Results for Synthetic Directed Networks with Simulated Edge Probabilities} \label{results_synthetic_networks}

In Table \ref{table:pearson_synthetic_ISM_standard} and Table \ref{table:spearman_synthetic_ISM_standard}, we have computed Pearson and Spearman's rank correlations, respectively, between ISM and standard centrality measures for all possible combinations of synthetic networks and edge probability distributions. These two tables reveal the impact of the edge probability increasing on average for the selected synthetic networks, since the used beta distributions have increasing expected values. Additionally, when going from left to right in the table, for each synthetic network type there is a general pattern of increasing number of edges. With an increasing number of edges, there comes increasing connectivity, which allows more probabilistic influence spread through a network.

From Table \ref{table:spearman_synthetic_ISM_standard}, we see that for all combinations of synthetic networks and edge probability, out-centrality has a strong positive Spearman's rank correlation with weighted out-degree. Combined with the Pearson correlation results in Table \ref{table:pearson_synthetic_ISM_standard}, we can infer that for some networks the relationship between out-centrality and weighted out-degree can be linear, but for most networks it is non-linear and monotonic. The ISM betweenness centrality and standard betweenness centrality have Pearson correlations ranging from strong to moderate to near zero values. Spearman's rank correlation values hint at a positive non-linear association of varying strength between these centrality measures. These two betweenness centrality measures indeed measure different things, and the empirical results support this notion. Spearman's rank correlation varies from strong to moderately positive between out-centrality and outward closeness centrality for all synthetic network and edge probability combinations. Thus, a possible non-linear and monotonic relationship between the measures is possible, although this relationship is clearly weaker than the relationship between out-centrality and weighted out-degree.

The Pearson correlations between in-centrality and weighted in-degree and between in-centrality and weighted inward closeness display correlations ranging from strongly or moderately positive to very weak near-zero values. Network connectivity has an effect on in-centrality values by causing reversing ranking behavior in in-centrality. Spearman's rank correlations range from strongly positive to strongly/moderately negative between in-centrality and weighted in-degree and between in-centrality and weighted inward closeness centrality. The relationship between in-centrality and out-centrality is either non-existent or has a strongly negative association in the synthetic networks. Both Pearson and Spearman's rank correlations vary from values close to zero to strongly negative correlation. 

For out-centrality and outward Katz centrality, we see a strong or moderately positive Pearson correlation becoming weaker as edge probability increases. This finding is consistent with the findings in  \cite{luhtala2025} for undirected networks. When edge probabilities are higher on average, out-centrality values tend to have higher values closer to one, which causes a non-linear relationship even with normalized Katz centrality values. The non-linear yet monotonic relationship between out-centrality and Katz centrality can be seen in Table \ref{table:spearman_synthetic_ISM_standard}, where Spearman's rank correlations are strongly positive for all combinations of synthetic network and edge probability distributions.

For in-centrality and inward Katz centrality, the story is different compared to the relationship between out-centrality and outward Katz centrality. In Table \ref{table:pearson_synthetic_ISM_standard}, we see that all the scale-free networks display Pearson correlations near zero. For directed Erd\H{o}s-R\'enyi networks with lower connectivity (lower values of edge creation probability) and edge probability distributions with lower expected values, there is a stronger positive Pearson correlation. As network connectivity and the average edge probability increase, the Pearson correlation approaches zero. Navigable small-world networks display similar phenomena: When connectivity increases and average edge probability increases, Pearson correlation approaches zero values. In Table \ref{table:spearman_synthetic_ISM_standard}, we see that Spearman's rank correlations vary from  strongly positive to strongly negative between in-centrality and inward Katz centrality. The inverting ranking behaviour of in-centrality likely plays a major role in this phenomenon.

\begin{table}
\vspace{-6pt}
\caption{Pearson correlation coefficients between ISM centrality measures and standard centrality measures for synthetic networks with simulated edge probabilities (wtd denotes weighted). The distributions used to simulate edge probabilities are abbreviated with Beta1 = $\text{Beta}(0.5, 4.5)$, Beta2 = $\text{Beta}(2, 8)$ and Beta3 = $\text{Beta}(1, 7/3)$.}
\begin{tabular}{|p{1.70cm}|p{0.85cm}|p{0.95cm}|p{0.95cm}|p{0.95cm}|p{0.95cm}|p{0.95cm}|p{0.95cm}|p{0.95cm}|p{0.95cm}|p{0.95cm}|}
 \hline
 \multicolumn{2}{|c|}{} &
 \multicolumn{9}{|c|}{Synthetic Network} \\
 \hline
 Comparison & Distr. & DER1 & DER2 & DER3 & NSW1 & NSW2 & NSW3 & DSF1 & DSF2 & DSF3 \\
 \hline
 \multirow{3}{\linewidth}{Out-c. VS wtd. out-degree} & Beta1 & 0.933 & 0.929 & 0.704 & 0.864 & 0.933 & 0.880 & 0.696 & 0.470 & 0.430  \\
 & Beta2 & 0.887 & 0.828 & 0.480 & 0.930 & 0.888 & 0.850 & 0.556 & 0.359 & 0.325 \\
 & Beta3 & 0.859 & 0.553 & 0.160 & 0.859 & 0.625 & 0.449 & 0.459 & 0.286 & 0.249 \\
 \hline
 \multirow{3}{\linewidth}{out-c. VS wtd. outw. closeness} & Beta1 & 0.800 & 0.982 & 0.683 & 0.789 & 0.881 & 0.843 & 0.780 & 0.836 & 0.870 \\
 & Beta2 & 0.867 & 0.841 & 0.446 & 0.815 & 0.751 & 0.782 & 0.855 & 0.884 & 0.862 \\
 & Beta3 & 0.920 & 0.558 & 0.135 & 0.745 & 0.544 & 0.348 & 0.924 & 0.920 & 0.850 \\
 \hline
 \multirow{3}{\linewidth}{ISM btw. VS wtd. std. btw.} & Beta1 & 0.917 & 0.968 & 0.515 & 0.817 & 0.838 & 0.728 & 0.867 & 0.936 & 0.908 \\
 & Beta2 & 0.966 & 0.700 & 0.290 & 0.840 & 0.617 & 0.619 & 0.811 & 0.935 & 0.911 \\
 & Beta3 & 0.863 & 0.337 & 0.062 & 0.551 & 0.228 & 0.218 & 0.720 & 0.881 & 0.843 \\
 \hline
 \multirow{3}{\linewidth}{in-c. VS wtd. in-degree} & Beta1 & 0.929 & 0.729 & -0.043 & 0.861 & 0.606 & 0.010 & 0.077 & 0.026 & 0.038 \\
 & Beta2 & 0.880 & 0.121 & 0.015 & 0.607 & -0.087 & -0.011 & 0.079 & 0.027 & 0.039 \\
 & Beta3 & 0.290 & 0.063 & -0.032 & 0.037 & -0.056 & 0.003 & 0.079 & 0.029 & 0.040 \\
 \hline
 \multirow{3}{\linewidth}{in-c. VS wtd. inw. closeness} & Beta1 & 0.763 & 0.873 & -0.048 & 0.846 & 0.764 & 0.011 & 0.510 & 0.280 & 0.332 \\
 & Beta2 & 0.851 & 0.124 & 0.021 & 0.866 & -0.143 & -0.027 & 0.576 & 0.324 & 0.409 \\
 & Beta3 & 0.654 & 0.049 & -0.008 & -0.015 & -0.039 & 0.046 & 0.450 & 0.236 & 0.286 \\
 \hline
 \multirow{3}{\linewidth}{in-c. VS out-c.} & Beta1 & -0.043 & -0.075 & -1.000 & 0.017 & -0.156 & -0.997 & 0.086 & 0.027 & 0.109 \\
 & Beta2 & -0.031 & -0.995 & -1.000 & 0.166 & -0.999 & -1.000 & 0.114 & 0.051 & 0.093 \\
 & Beta3 & -0.030 & -0.999 & -1.000 & -0.979 & -1.000 & -1.000 & 0.105 & 0.047 & 0.099 \\
 \hline
 \multirow{3}{\linewidth}{out-c. VS outw. Katz (0.9)} & Beta1 & 0.941 & 0.969 & 0.688 & 0.954 & 0.913 & 0.860 & 0.752 & 0.675 & 0.595 \\
 & Beta2 & 0.999 & 0.804 & 0.472 & 0.950 & 0.828 & 0.838 & 0.597 & 0.541 & 0.457 \\
 & Beta3 & 0.873 & 0.529 & 0.159 & 0.757 & 0.571 & 0.435 & 0.496 & 0.441 & 0.357 \\
 \hline
 \multirow{3}{\linewidth}{out-c. VS outw. Katz (0.5)} & Beta1 & 1.000 & 0.967 & 0.699 & 0.963 & 0.943 & 0.876 & 0.737 & 0.602 & 0.536 \\
 & Beta2 & 0.969 & 0.823 & 0.477 & 0.974 & 0.877 & 0.847 & 0.584 & 0.475 & 0.409 \\
 & Beta3 & 0.896 & 0.546 & 0.159 & 0.849 & 0.610 & 0.444 & 0.486 & 0.386 & 0.319 \\
 \hline
 \multirow{3}{\linewidth}{out-c. VS outw. Katz (0.1)} & Beta1 & 0.954 & 0.940 & 0.704 & 0.886 & 0.938 & 0.880 & 0.706 & 0.500 & 0.454  \\
 & Beta2 & 0.906 & 0.829 & 0.480 & 0.942 & 0.888 & 0.850 & 0.563 & 0.385 & 0.343 \\
 & Beta3 & 0.871 & 0.553 & 0.160 & 0.861 & 0.624 & 0.449 & 0.466 & 0.308 & 0.265 \\
 \hline
 \multirow{3}{\linewidth}{in-c. VS inw. Katz (0.9)} & Beta1 & 0.925 & 0.768 & -0.049 & 0.785 & 0.649 & 0.011 & 0.078 & 0.039 & 0.055 \\
 & Beta2 & 0.992 & 0.123 & 0.007 & 0.735 & -0.166 & -0.019 & 0.081 & 0.043 & 0.057 \\
 & Beta3 & 0.277 & 0.060 & -0.037 & -0.055 & -0.079 & 0.005 & 0.082 & 0.044 & 0.059 \\
 \hline
 \multirow{3}{\linewidth}{in-c. VS inw. Katz (0.5)} & Beta1 & 1.000 & 0.763 & -0.047 & 0.919 & 0.638 & 0.010 & 0.079 & 0.035 & 0.049 \\
 & Beta2 & 0.962 & 0.123 & 0.011 & 0.678 & -0.122 & -0.015 & 0.081 & 0.038 & 0.051 \\
 & Beta3 & 0.290 & 0.061 & -0.035 & -0.007 & -0.070 & 0.005 & 0.082 & 0.039 & 0.052 \\
 \hline
 \multirow{3}{\linewidth}{in-c. VS inw. Katz (0.1)} & Beta1 & 0.951 & 0.738 & -0.044 & 0.877 & 0.613 & 0.010 & 0.078 & 0.028 & 0.040 \\
 & Beta2 & 0.900 & 0.122 & 0.014 & 0.621 & -0.093 & -0.012 & 0.079 & 0.030 & 0.042 \\
 & Beta3 & 0.291 & 0.062 & -0.032 & 0.029 & -0.058 & 0.003 & 0.080 & 0.031 & 0.043 \\
 \hline
\end{tabular}
\label{table:pearson_synthetic_ISM_standard}
\end{table}

\begin{table}
\vspace{-6pt}
\caption{Spearman's rank correlation coefficients between ISM centrality measures and standard centrality measures for synthetic networks with simulated edge probabilities (wtd denotes weighted). The distributions used to simulate edge probabilities are abbreviated with Beta1 = $\text{Beta}(0.5, 4.5)$, Beta2 = $\text{Beta}(2, 8)$ and Beta3 = $\text{Beta}(1, 7/3)$.}
\begin{tabular}{|p{1.70cm}|p{0.85cm}|p{0.95cm}|p{0.95cm}|p{0.95cm}|p{0.95cm}|p{0.95cm}|p{0.95cm}|p{0.95cm}|p{0.95cm}|p{0.95cm}|}
 \hline
 \multicolumn{2}{|c|}{} &
 \multicolumn{9}{|c|}{Synthetic Network} \\
 \hline
 Comparison & Distr. & DER1 & DER2 & DER3 & NSW1 & NSW2 & NSW3 & DSF1 & DSF2 & DSF3 \\
 \hline
 \multirow{3}{\linewidth}{Out-c. VS wtd. out-degree} & Beta1 & 0.948 & 0.945 & 0.990 & 0.880 & 0.969 & 0.986 & 0.969 & 0.993 & 0.997 \\
 & Beta2 & 0.883 & 0.992 & 0.994 & 0.940 & 0.992 & 0.991 & 0.988 & 0.997 & 0.999 \\
 & Beta3 & 0.918 & 0.971 & 0.969 & 0.961 & 0.956 & 0.954 & 0.987 & 0.994 & 0.996 \\
 \hline
 \multirow{3}{\linewidth}{out-c. VS wtd. outw. closeness} & Beta1 & 0.928 & 0.981 & 0.872 & 0.843 & 0.881 & 0.905 & 0.969 & 0.955 & 0.915 \\
 & Beta2 & 0.996 & 0.931 & 0.839 & 0.790 & 0.814 & 0.897 & 0.946 & 0.922 & 0.912 \\
 & Beta3 & 0.942 & 0.820 & 0.682 & 0.787 & 0.769 & 0.791 & 0.939 & 0.893 & 0.856 \\  
 \hline
 \multirow{3}{\linewidth}{ISM btw. VS wtd. std. btw.} & Beta1 & 0.904 & 0.973 & 0.614 & 0.832 & 0.855 & 0.767 & 0.867 & 0.857 & 0.893 \\
 & Beta2 & 0.963 & 0.760 & 0.387 & 0.834 & 0.670 & 0.637 & 0.899 & 0.877 & 0.884 \\
 & Beta3 & 0.929 & 0.485 & 0.175 & 0.660 & 0.370 & 0.307 & 0.866 & 0.814 & 0.813 \\
 \hline
 \multirow{3}{\linewidth}{in-c. VS wtd. in-degree} & Beta1 & 0.949 & 0.918 & -0.027 & 0.870 & 0.774 & -0.026 & -0.298 & -0.715 & -0.841 \\
 & Beta2 & 0.873 & 0.085 & 0.031 & 0.684 & -0.100 & 0.002 & -0.619 & -0.823 & -0.891 \\
 & Beta3 & 0.852 & 0.074 & 0.033 & 0.027 & -0.080 & -0.034 & -0.627 & -0.803 & -0.879 \\
 \hline
 \multirow{3}{\linewidth}{in-c. VS wtd. inw. closeness} & Beta1 & 0.922 & 0.986 & -0.055 & 0.863 & 0.870 & -0.004 & -0.308 & -0.611 & -0.718 \\
 & Beta2 & 0.996 & 0.076 & 0.022 & 0.964 & -0.127 & -0.018 & -0.558 & -0.717 & -0.759 \\
 & Beta3 & 0.948 & 0.065 & 0.065 & 0.009 & -0.087 & -0.028 & -0.494 & -0.635 & -0.667 \\
 \hline
 \multirow{3}{\linewidth}{in-c. VS out-c.} & Beta1 & -0.048 & -0.119 & -1.000 & 0.013 & -0.315 & -0.998 & -0.791 & -0.973 & -0.975 \\
 & Beta2 & -0.010 & -0.997 & -1.000 & 0.149 & -0.999 & -1.000 & -0.918 & -0.983 & -0.977 \\
 & Beta3 & -0.130 & -1.000 & -0.989 & -0.989 & -1.000 & -1.000 & -0.932 & -0.983 & -0.976 \\
 \hline
 \multirow{3}{\linewidth}{out-c. VS outw. Katz (0.9)} & Beta1 & 0.955 & 0.994 & 0.973 & 0.981 & 0.973 & 0.965 & 0.979 & 0.978 & 0.981 \\
 & Beta2 & 0.999 & 0.972 & 0.983 & 0.980 & 0.925 & 0.975 & 0.969 & 0.981 & 0.989 \\
 & Beta3 & 0.969 & 0.941 & 0.959 & 0.905 & 0.878 & 0.933 & 0.963 & 0.973 & 0.984 \\
 \hline
 \multirow{3}{\linewidth}{out-c. VS outw. Katz (0.5)} & Beta1 & 1.000 & 0.989 & 0.986 & 0.966 & 0.993 & 0.983 & 0.984 & 0.985 & 0.988  \\
 & Beta2 & 0.967 & 0.990 & 0.991 & 0.989 & 0.981 & 0.986 & 0.987 & 0.986 & 0.993 \\
 & Beta3 & 0.975 & 0.965 & 0.966 & 0.969 & 0.937 & 0.947 & 0.982 & 0.981 & 0.989 \\
 \hline
 \multirow{3}{\linewidth}{out-c. VS outw. Katz (0.1)} & Beta1 & 0.965 & 0.956 & 0.991 & 0.900 & 0.977 & 0.987 & 0.975 & 0.994 & 0.997 \\
 & Beta2 & 0.902 & 0.994 & 0.994 & 0.952 & 0.992 & 0.991 & 0.991 & 0.997 & 0.999 \\
 & Beta3 & 0.933 & 0.972 & 0.969 & 0.966 & 0.955 & 0.953 & 0.989 & 0.994 & 0.996 \\
 \hline
 \multirow{3}{\linewidth}{in-c. VS inw. Katz (0.9)} & Beta1 & 0.953 & 0.998 & -0.034 & 0.955 & 0.865 & -0.019 & -0.361 & -0.715 & -0.826 \\
 & Beta2 & 0.999 & 0.077 & 0.029 & 0.919 & -0.174 & -0.012 & -0.621 & -0.822 & -0.887 \\
 & Beta3 & 0.978 & 0.082 & 0.041 & -0.052 & -0.126 & -0.041 & -0.621 & -0.797 & -0.869 \\
 \hline
 \multirow{3}{\linewidth}{in-c. VS inw. Katz (0.5)} & Beta1 & 1.000 & 0.976 & -0.031 & 0.939 & 0.829 & -0.022 & -0.336 & -0.716 & -0.832 \\
 & Beta2 & 0.964 & 0.081 & 0.030 & 0.790 & -0.132 & -0.005 & -0.627 & -0.825 & -0.889 \\
 & Beta3 & 0.939 & 0.079 & 0.038 & -0.009 & -0.104 & -0.037 & -0.631 & -0.801 & -0.874 \\
 \hline
 \multirow{3}{\linewidth}{in-c. VS inw. Katz (0.1)} & Beta1 & 0.966 & 0.932 & -0.028 & 0.886 & 0.786 & -0.025 & -0.307 & -0.718 & -0.840 \\
 & Beta2 & 0.895 & 0.085 & 0.031 & 0.703 & -0.105 & 0.001 & -0.623 & -0.826 & -0.892 \\
 & Beta3 & 0.872 & 0.074 & 0.034 & 0.021 & -0.084 & -0.035 & -0.631 & -0.805 & -0.879 \\
 \hline
\end{tabular}
\label{table:spearman_synthetic_ISM_standard}
\end{table}

\section{Discussion} \label{discussion}

The main objective of this study is to clarify when standard directed and weighted centrality measures can effectively serve as proxies for probabilistic influence, and when explicitly probabilistic ISM centrality measures offer additional insights. Our findings indicate that this depends on the type of node role being measured, the network structure, and the distribution of edge probabilities.

For broadcasting influence, we observe a fairly consistent relationship between ISM out-centrality and standard centrality measures. Specifically, out-centrality demonstrates strong to moderate positive Pearson correlations with weighted out-degree and outward Katz centrality across the empirical and synthetic networks studied, while Spearman's rank correlation is strong in all the networks studied. This suggests that standard measures can often be useful proxies for identifying influential source nodes. The Pearson correlation is particularly robust when edge weights are low, as influence propagation tends to be local, with shorter walks dominating the transmission probabilities. However, as edge probabilities increase, influence can spread through longer paths, leading to more variable Pearson correlations. This variation highlights the fact that standard measures may not fully account for the probabilistic dependencies and overlaps between multiple influence pathways.

The situation regarding the reception of influence is more complex than initially appears. The relationship between ISM in-centrality and other receiving-oriented measures, such as weighted in-degree, inward closeness, and inward Katz centrality, largely depends on the network structure and edge probabilities. This finding supports the idea that susceptibility to influence is not dictated solely by local incoming connections; rather, it emerges from the broader context of the network \cite{luhtala2025}. Notably, the ranking reversal phenomenon of in-centrality in highly connected networks suggests that nodes that are weakly influenced when edge probabilities are low can become strongly influenced when transmission probabilities increase. Therefore, standard inward centrality measures may only approximate probabilistic susceptibility under certain network conditions and edge weight scenarios.

The ISM betweenness centrality also differs conceptually from the shortest-path betweenness. Although the standard betweenness centrality is based on shortest paths, ISM betweenness centrality assesses how the total amount of probabilistic influence in the network changes when a node is removed. It accounts for finite walks, alternative routes, redundant paths, and circular propagation. This distinction clarifies why the ISM betweenness and the shortest-path betweenness may identify different types of intermediary nodes. In networks where influence can propagate through multiple overlapping pathways, the shortest-path betweenness might overlook nodes that are crucial for probabilistic transmission, even if they do not lie on many shortest paths.

One significant advantage of the ISM framework is its probabilistic interpretation. Although weighted standard centrality measures include edge weights, their values typically do not have a direct interpretation as probabilities of influence or transmission. In contrast, the influence-spreading model is explicitly built on the probabilities of edge transmission and analytically combines these walks of finite length. As a result, the ISM out-centralities and in-centralities can be interpreted directly in terms of probabilistic influence, making them particularly valuable for applications such as influence maximization, targeted information dissemination, epidemic modeling, and misinformation mitigation.

The findings of the present study also underscore the importance of the magnitudes of edge probability. When edge probabilities are low, the influence-spreading model calculations converge quickly because longer walks contribute only minimally to the total transmission probability. This helps explain why correlations with standard measures are often stronger in this context: influence tends to be primarily local, and standard structural measures capture much of the relevant information. However, as edge probabilities increase, the significance of longer walks and multi-walk dependencies rises, leading to a divergence between the ISM centralities and standard centralities. Thus, the relationship between structural centrality and probabilistic diffusion is not fixed, as it strongly depends on the expected range of influence spreading.

From a computational point of view, ISM is manageable for large networks when implemented efficiently. With an optimized approach \cite{kuikka2022efficiency}, networks with up to approximately 100,000 nodes can be analyzed if the graph is sufficiently sparse. However, scalability is not just dependent on the number of nodes, but also on the number of edges, network density, intersections of walks, and the selected maximum walk length \(L_{max}\). Reducing \(L_{max}\) is a straightforward way to decrease computation time, though it may neglect the effects of longer-range influence. Future research should therefore explore scalable approximation methods, the impact of different \(L_{max}\) choices, and potential extensions of the influence-spreading model to temporal, multiplex, or multilayer networks.

\section{Conclusion} \label{conclusion}

This study compares the probabilistic centrality measures of the Influence-Spreading Model (ISM), specifically out-centrality, in-centrality, and ISM betweenness centrality with standard directed and weighted centrality measures. Using both real-world online social networks and synthetic networks, we assessed the Pearson and Spearman's rank correlation between these measures to understand when standard centrality approximates probabilistic influence.

The results show that commonly used centrality measures can be useful proxies for ISM centralities in certain contexts. In particular, ISM out-centrality values consistently correlated with weighted out-degree and outward Katz centrality values, especially when the edge probabilities were low. In the studied networks, rank correlation was found to be strong between the ISM out-centrality and weighted out-degree as well as outward Katz centrality. In contrast, the ISM in-centrality showed more variability, indicating that susceptibility to influence is more dependent on network structure. Furthermore, ISM betweenness centrality accounts for probabilistic transmission through multiple paths, unlike standard shortest-path betweenness.

In general, while standard measures can approximate influence in simple local contexts, ISM provides deeper insight when influence spreads through longer or overlapping pathways. Its centrality values, derived from a clear probabilistic model, offer interpretations that traditional measures often lack, making ISM a valuable framework for studying influence and information diffusion in directed and weighted networks.

\section*{Author contributions statement}
Juuso Luhtala: Conceptualization; Methodology; Investigation; Data curation; Formal analysis (interpretation of results); Writing--review \& editing. \\
Vesa Kuikka: Methodology (model utilization and formal modeling framework); Writing--review \& editing. \\
Kimmo Kaski: Supervision; Review \& editing.

All authors have read and approved the final manuscript.

\section*{Additional information}
\textbf{Competing interests}: The authors declare no competing interests \\
\textbf{Funding}: This research received no external funding. \\
\textbf{Data Availability}: The online social datasets analyzed in this study were derived from previously published sources, where the raw data and empirically estimated influence probabilities are publicly available. Specifically, the data can be obtained from the following sources: the Twitter influence dataset by Zhang \emph{et al.} (DOI: \href{https://doi.org/10.1145/3308560.3316701}{10.1145/3308560.3316701}); the congressional Twitter network dataset by Fink \emph{et al.} (DOI: \href{https://doi.org/10.1016/j.dib.2023.109521}{10.1016/j.dib.2023.109521}); and the content-based network dataset by Logins and Karras (DOI: \href{https://doi.org/10.1109/ICDMW.2019.00020}{10.1109/ICDMW.2019.00020}).

%
%


\bibliography{References_ISM_walks_comparison}

\end{document}